\documentclass[aps,prb,twocolumn,showpacs]{revtex4-1}%
\usepackage{epsfig}
\usepackage{color}
\usepackage{amsmath}
\usepackage{amssymb}
\usepackage{amsfonts}
\usepackage{bm}
\usepackage{dcolumn}
\usepackage{graphicx}
\usepackage{subfigure}
\usepackage{natbib}
\usepackage{float}%
\providecommand{\U}[1]{\protect\rule{.1in}{.1in}}

\begin{document}
\title{Charge Transport in Hybrid Halide Perovskites}
\author{Mingliang Zhang$^{1,2}$, Xu Zhang$^{2}$, Ling-Yi Huang$^{2}$, Hai-Qing
Lin$^{1\dagger}$ and Gang Lu$^{2\ast}$}
\affiliation{$^{1}$Beijing Computational Science Research Center, Beijing 100193, China}
\affiliation{$^{2}$Department of Physics and Astronomy, California State University
Northridge, Northridge, CA 91330, USA}


\begin{abstract}
Charge transport is crucial to the performance of hybrid halide perovskite
solar cells. A theoretical model based on large polarons is proposed to
elucidate charge transport properties in the perovskites. Critical new
physical insights are incorporated into the model, including the recognitions
that acoustic phonons as opposed to optical phonons are responsible for the
scattering of the polarons; these acoustic phonons are fully excited due to
the ``softness'' of the perovskites, and the temperature-dependent dielectric
function underlies the temperature dependence of charge carrier mobility. This
work resolves key controversies in literature and forms a starting point for
more rigorous first-principles predictions of charge transport.

\end{abstract}
\maketitle



\section{introduction}

Organic-inorganic hybrid perovskites represent a fascinating class of
materials poised to revolutionize optoelectronic, in particular, photovoltaic
applications \cite{gra,don,sai}. These materials possess a set of unusual
transport properties crucial to their photovoltaic performance. Essential to
the transport properties is charge carrier mobility $\mu$, which exhibits
following behavior unique to this family of materials: (1)$\mu\varpropto
n^{-1}$ where $n$ is charge carrier concentration \cite{bi}; (2)$\mu\varpropto
I_{0}^{-1/2}$ where $I_{0}$ is incident photon flux \cite{oga}; (3)$\mu
\varpropto T^{-3/2}$ where $T$ is temperature \cite{mil,sav,oga,kar,hut}; and
(4)$\mu$ is insensitive to defects \cite{zhu,bre}. There is great interest to
understand and control the transport properties of the perovskites, further
propelling the development of perovskite-based solar cells. However, no
complete physical picture has emerged so far to fully account for the
experimental observations on charge transport and the nature of charge
transport remains a subject of intense debate \cite{zhu,bre,wri,sen}.

In this paper, we propose a theoretical model to elucidate the charge
transport behavior in the perovskites. In this model, the charge carriers are
characterized as large polarons, resulted from the carrier interaction with
optical phonons \cite{zhu}. Hence the residual interaction between the
polarons and the optical phonons is much weaker than the interaction with
acoustic phonons. The charge transport is determined by the scattering of the
polarons by themselves, defects and longitudinal acoustic (LA) phonons and is
governed by Boltzmann equation. These interactions are screened by a
temperature-dependent dielectric function as a result of spontaneous
polarization in the perovskites at low temperatures. Owing to the
\textquotedblleft softness\textquotedblright\ of the perovskites, the LA
phonons are fully excited, interacting strongly with the polarons. The
constant carrier concentration $n$ leads to an equilibrium distribution
function of the polarons that is proportional to $n$, resulting in
$n$-dependent carrier mobilities.

\section{Nature of charge carriers}

In the following, we take MAPbI$_{3}$ [MA$^{+}$=(CH$_{3}$NH$_{3}$)$^{+}$] as a
representative of ABX$_{3}$ perovskite family to illustrate the general
physical picture of charge transport.

\subsection{Properties of of large polarons}

In MAPbI$_{3}$, the interaction between a free carrier and longitudinal
optical (LO) phonons (i.e., Pb-I stretching modes) is stronger than that
between the carrier and the acoustic phonons \cite{cal}, supported by the
emission line broadening experiment \cite{wri}. According to the theory of
large polarons \cite{emin}, the binding energy, radius and effective mass of a
large polaron can be expressed as $E_{\text{P}}=V_{L}^{2}/4T_{e}$,
$R_{\text{P}}=2T_{e}a/V_{L}$, and $m_{\text{P}}=V_{L}^{4}[4\omega_{\text{LO}%
}^{2}a^{2}T_{e}^{3}]^{-1}$, respectively. Here $\omega_{\text{LO}}$ is the
frequency of the LO phonon and $a$ is the lattice constant of MAPbI$_{3}$;
$T_{e}\thicksim\hbar^{2}/(2mr^{2})$ is the kinetic energy of the conduction
electrons, where $m$ is the mass of the electron, $r$ is the characteristic
length-scale over which the wave-function of the conduction electron changes
substantially, taken as the mean value between the radius of Pb$^{2+}$ ion and
Pb atom \cite{ii,ir}, i.e., $r\thicksim$1.675 \AA . $V_{L}$ represents the
interaction between the carrier and the LO phonon-induced electric field,
\begin{equation}
V_{L}\thicksim\frac{1}{4\pi\epsilon_{0}}\frac{e^{2}}{2}(\frac{1}%
{\varepsilon_{\infty}}-\frac{1}{\varepsilon_{0}})\frac{1}{r}, \label{pot}%
\end{equation}
where $\varepsilon_{0}$ and $\varepsilon_{\infty}$ are static and optical
dielectric constant. Using both experimentally measured \cite{lin,ono} and
first-principles computed \cite{fro14,bri} parameters of MAPbI$_{3}$, we
estimate $E_{\text{P}}\thicksim$ 67 - 112 meV, $R_{\text{P}}\thicksim$ 22 - 28
\AA \ and $m_{\text{P }}\thicksim$ $4.1-12$ $m$. Since $E_{\text{P}}$ is much
higher than the room temperature, these polarons are thermally stable, in line
with the large polaron hypothesis for charge transport
\cite{zhu,bre,sen,pro,fro14,bok,ost}. We can also estimate the critical
concentration of the polarons as $n_{c}=(2R_{\text{P}})^{-3}\thicksim
5.5\times10^{18}$cm$^{-3}$; beyond this critical value, neighboring polarons
would overlap. In normal operating conditions of the solar cells, the free
carrier concentration $n$ is \ \cite{hut,mil} less than 10$^{18}$cm$^{-3}$ and
$n_{c}$, thus the large polarons could avoid each other in MAPbI$_{3}$.

\subsection{Distribution function of polarons}

\label{snd}

The Fermi-Dirac distribution of polarons can be approximated by the Boltzmann
distribution \cite{v5} if
\begin{equation}
(\frac{2\pi\hbar^{2}}{m_{\text{P}}k_{B}T})^{3/2}n\ll1. \label{bd}%
\end{equation}
Under the normal operating conditions, this equation is satisfied, thus the
photo-generated electrons are non-degenerate and one can replace the
Fermi-Dirac distribution by the Boltzmann distribution. Later we will show
that the polaron state can be characterized by its momentum $\mathbf{p}$, and
the energy of the polaron state $|\mathbf{p}\rangle$ is thus denoted as
$\varepsilon_{\mathbf{p}}$.

In an intrinsic or lightly doped MAPbI$_{3}$, the carriers are generated
primarily by photo- as opposed to thermal excitations. Thus $n$ is determined
by $I_{0}$, and largely independent \cite{oga,che} of $T$. Hence, we can
express the occupation number $f_{0}(\varepsilon_{\mathbf{p}})$ per spin for
polarons of energy $\varepsilon_{\mathbf{p}}$ as:%
\begin{equation}
f_{0}(\varepsilon_{\mathbf{p}})=\frac{4\pi^{3/2}\hbar^{3}e^{-\varepsilon
_{\mathbf{p}}/k_{B}T}}{(2m_{\text{P}}k_{B}T)^{3/2}}n. \label{fd}%
\end{equation}
As will be shown later, the linear $n$ dependence of the polaron distribution
function gives rise to the $n$ dependence of carrier mobility.

\subsection{Formation free energy of polarons}

To further establish the fact that the polarons are thermodynamically stable
than free electrons in MAPbI$_{3}$ under the normal operating conditions, we
next estimate the formation \textit{free} energy of the polarons relative to
that of the free electrons. There are two major contributions to the entropy
of the polarons. Once a polaron is formed, it acquires an excluded volume and
increases its effective mass, leading to higher translational entropy. At the
same time, the induced lattice distortion due to the polaron increases the
vibrational frequencies and lowers the vibrational entropy. In the following,
we will estimate these competing contributions to the entropy.

\subsubsection{Change in translational entropy}

It is known that the translational entropy for non-degenerate free electron
gas of $N$ electrons occupying a volume of $V$, is
\begin{equation}
S_{\text{g}}=Nk_{B}[\ln\frac{V}{N}(\frac{2\pi mk_{B}T}{h^{2}})^{3/2}+\frac
{5}{2}], \label{gne}%
\end{equation}
where $h$ is the Planck constant.\cite{hill}

If all electrons become polarons, the free volume $V_{f}$ for the polarons is
reduced to $V_{f}=V-N4\pi R_{\text{P}}^{3}/3$, and their corresponding
translational entropy $S_{\text{P}}$ becomes \cite{hill}%
\begin{equation}
S_{\text{P}}=Nk_{B}[\ln\frac{V_{f}}{N}(\frac{2\pi m_{\text{P}}k_{B}T}{h^{2}%
})^{3/2}+\frac{5}{2}], \label{gp}%
\end{equation}
where $m_{\text{P}}$ is the polaron mass. Thus, the change in entropy $\Delta
s_{e}$ per electron is
\begin{equation}
\Delta s_{e}=(S_{\text{P}}-S_{\text{g}})/N=k_{B}\ln(1-n\frac{4\pi R_{\text{P}%
}^{3}}{3})(\frac{m_{\text{P}}}{m})^{3/2}, \label{cen}%
\end{equation}
where $n=N/V$ is the number of free electrons per unit volume. From Eq.
(\ref{cen}), one can see that (i) $\Delta s_{e}$ does not depend on
temperature; (ii) the larger the $m_{\text{P}}$, the higher the entropy; (iii)
the finite size of the polarons decreases their entropy relative to the
electrons. Under the normal conditions, the concentration $n$ of the polarons
is much less than $n_{c}$, therefore the translational entropy of the polarons
is higher than that of the free electrons, i.e., $\Delta s_{e}>0$.

\subsubsection{Change in vibrational entropy}

The entropy $S_{1}$ of a harmonic oscillator with frequency $\omega$ is given
by \cite{v5}:
\begin{equation}
S_{1}=-k_{B}\ln(1-e^{-\hbar\omega/k_{B}T})+\frac{\hbar\omega}{T}%
\frac{e^{-\hbar\omega/k_{B}T}}{1-e^{-\hbar\omega/k_{B}T}}. \label{sz}%
\end{equation}
In a undeformed crystal, the entropy $S_{10}$ due to a single LO mode can be
obtained from Eq. (\ref{sz}) by letting $\omega=\omega_{\text{LO}}$, where
$\omega_{\text{LO}}$ is the frequency of the LO mode. In each primitive cell
of MAPbI$_{3}$, the vibrational frequencies of three stretching modes (Pb-I
bonds) are increased due to the lattice distortion. Hence, the vibrational
entropy is decreased.

\begin{figure}
[ht]\centering
\subfigure[]{\includegraphics[width=0.23\textwidth]{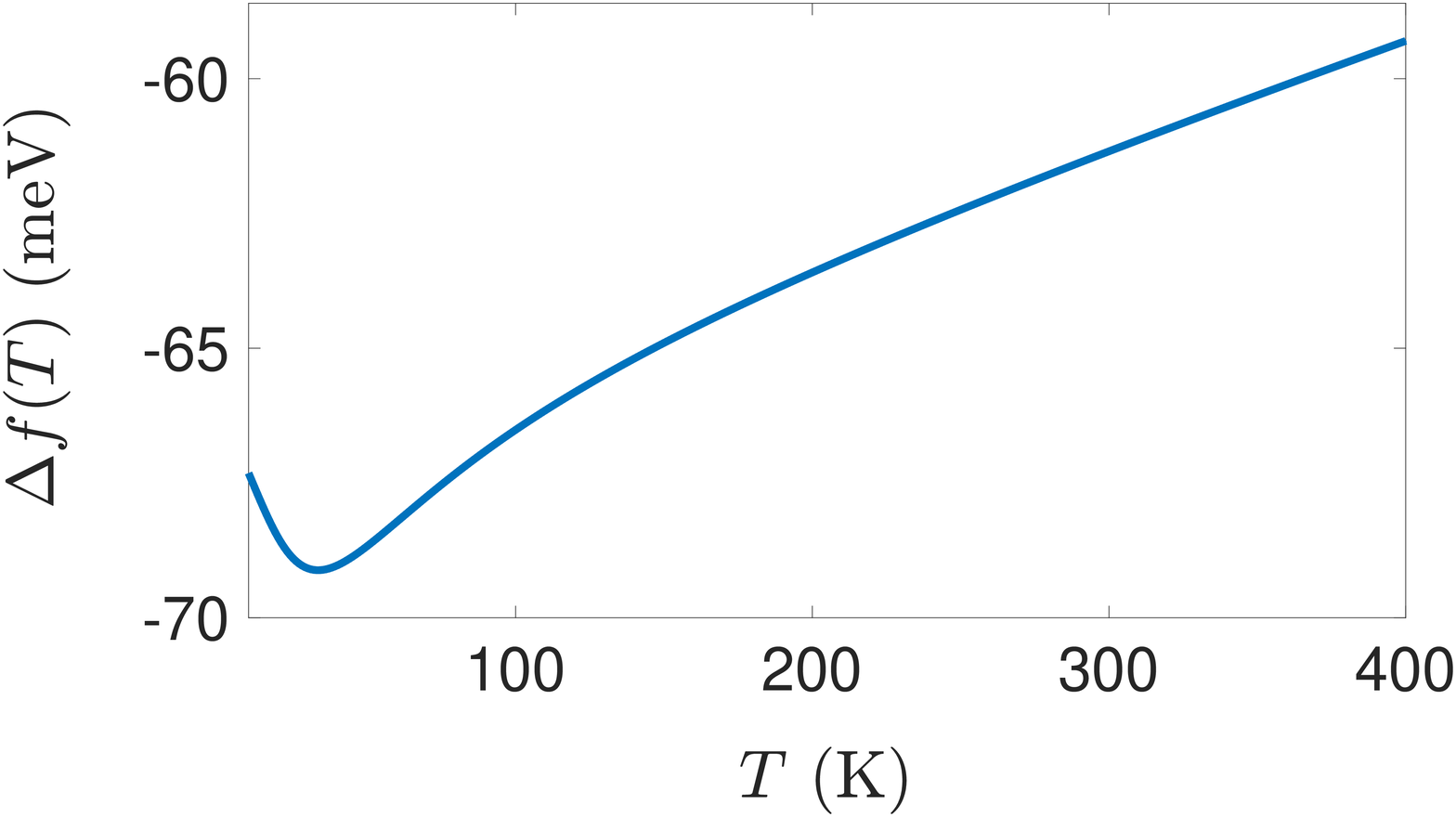}\label{frepol}}
\subfigure[]{\includegraphics[width=0.23\textwidth]{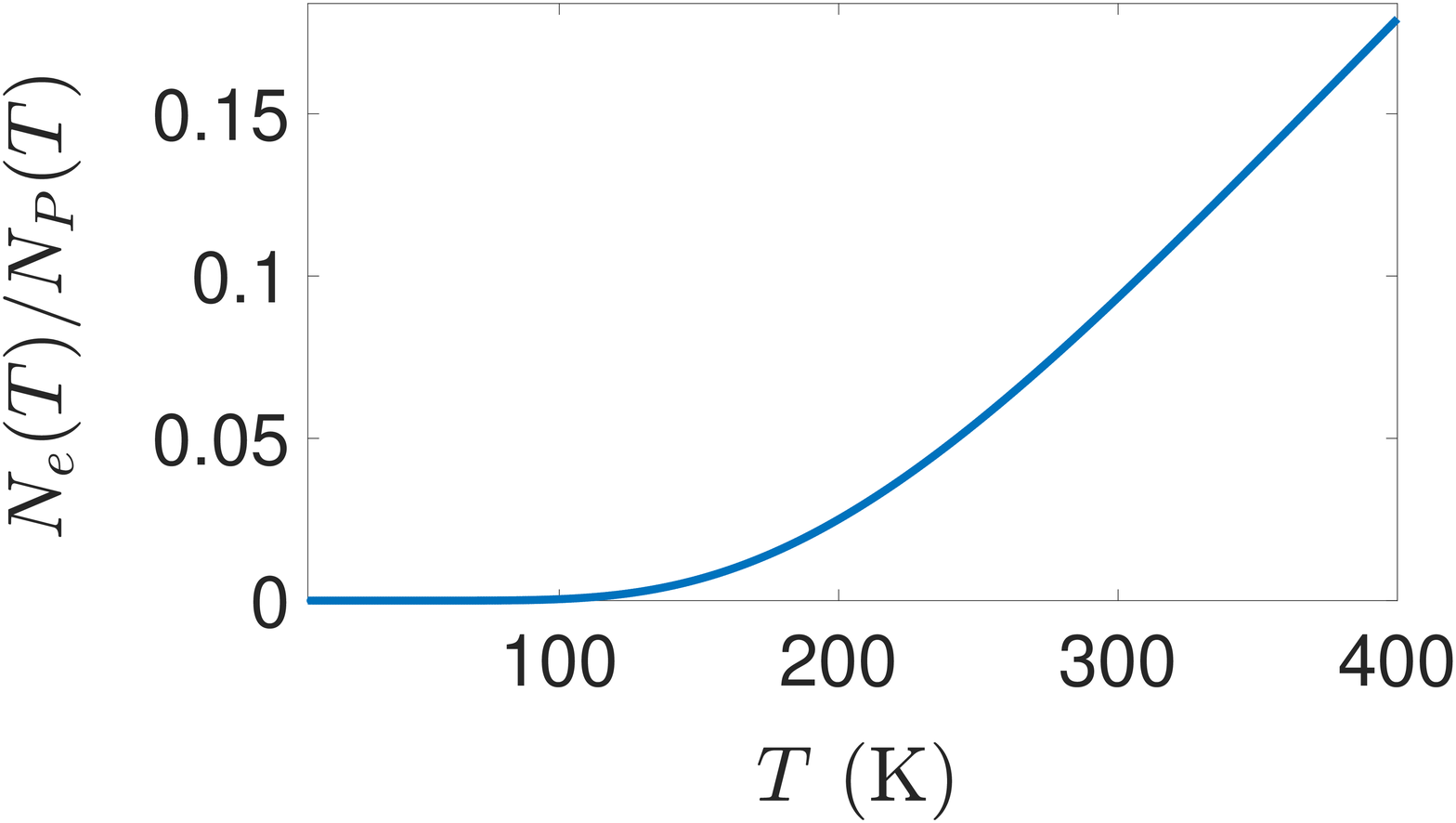}\label{Pfrac}}
\caption{(a) Formation free energy as function of temperature; (b) The ratio
of number of electrons ($N_{e}$) to the number of polarons ($N_{P}$) vs.
temperature.}
\end{figure}

Comparing to free electrons in a undeformed lattice, the change in the
vibrational entropy $\Delta s_{v}$ per polaron is:
\begin{equation}
\Delta s_{v}=-3\frac{\frac{4}{3}\pi R_{\text{P}}^{3}}{a^{3}}(S_{1}-S_{10}),
\label{ven}%
\end{equation}
where $\frac{4}{3}\pi R_{\text{P}}^{3}/a^{3}$ accounts for the number of the
primitive cells occupied by a large polaron, and the factor of 3 represents
the three Pb-I stretching modes. If temperature is higher than 50 K, the
decrease of the vibrational entropy dominates the change in the translational
entropy, cf. Fig.\ref{frepol}.

\subsubsection{Relative contributions to conductivity from polarons and
electrons}

The change in entropy $\Delta s$ in forming a polaron is
\begin{equation}
\Delta s=\Delta s_{e}+\Delta s_{v}. \label{tse}%
\end{equation}
The formation free energy $\Delta f$ per polaron is thus
\begin{equation}
\Delta f(T)=-E_{\text{P}}-T\Delta s, \label{df1}%
\end{equation}
which is plotted as a function of temperature in Fig.\ref{frepol}. The shallow
minimum in Fig.\ref{frepol} is due to the larger effective mass $m_{\text{P}}$
of polarons and $n\ll n_{c}$. With the increase of temperature, $|\Delta
f(T)|$ is decreased, i.e., polarons become less stable at a higher temperature.

At temperature $T$, the ratio between $N_{e}(T)$ (the number of electrons) and
$N_{\text{P}}(T)$ (the number of polarons) is given by \cite{v5}
\begin{equation}
\lbrack\frac{N_{e}(T)}{N_{\text{P}}(T)}]^{2}=e^{2\frac{\Delta f(T)}{k_{B}T}},
\label{ceq}%
\end{equation}
where we assume that the formation free energy of hole polarons is the same as
that of electron polarons. We can see from Fig.\ref{Pfrac} that the below 140
K, the number of electrons is negligible. At 300 K, $N_{e}(T)/N_{\text{P}%
}(T)\thicksim0.1$. Therefore, the dominant carriers in MAPbI$_{3}$ are large
polarons, as opposed to electrons and holes.

\section{dielectric screening}

There is a misconception in literature which attributes the temperature
dependence of carrier mobility, i.e., $\mu\propto T^{-3/2}$ entirely to the
scattering of acoustic phonons. This misconception counters the fact that many
non-polar semiconductors do not exhibit the same $T^{-3/2}$ dependence as the
perovskite materials although their carriers are scattered primarily by
acoustic phonons \cite{zeg}. We believe that the perovskites possess a unique
but often overlooked feature, i.e., the existence of a spontaneously polarized
phase at low temperatures, which is responsible for the unique temperature
dependence of carrier mobility. More specifically, we will reveal in following
that it is the temperature dependence of the dielectric function that among
other factors, yields the temperature dependence of carrier mobility in the perovskites.

Recent molecular dynamics simulations indicate that there exists a super
paraelectric phase in MAPbI$_{3}$ below 1000K [\onlinecite{fro14}]. It is
known that the super paraelectric phase emerges from a spontaneously polarized
phase with increased temperature. For ABX$_{3}$ perovskites, the critical
polarizability $\alpha_{c}$ above which a spontaneous polarization takes place
is given by $\alpha_{c}=(a/2)^{3}$/0.383 \ [\onlinecite{fey2}]. For
MAPbI$_{3}$, $\alpha_{c}=8.16\times10^{-29}$ m$^{3}$. On the other hand, the
polarizability of MAPbI$_{3}$, $\alpha_{\text{dis}}$ is mainly induced by the
displacements of Pb$^{2+}$ and I$^{-}$ ions and can be estimated as
$\alpha_{\text{dis}}=2.73\times10^{-28}$m$^{3}>\alpha_{c}$ [\onlinecite{sm}].
Hence, below a certain temperature, MAPbI$_{3}$ is spontaneously polarized.

For a super paraelectric phase, one can express its dielectric function as
follows: \cite{ger,pog,fey2,kits}%
\begin{equation}
\varepsilon(\omega,T)=\varepsilon_{\infty}+\frac{1}{3}\frac{n_{d}p^{2}}%
{k_{B}T\epsilon_{0}}\frac{1}{1-i\omega\tau(T)} \label{die0}%
\end{equation}%
\[
+\frac{9}{\beta(T-\theta)}\frac{\omega_{ip}^{2}}{\omega^{\prime2}%
-i\gamma^{\prime}\omega-\omega^{2}}.
\]
The first term represents the contribution from the bound electrons at the
optical frequencies and room temperature and it is taken from an experimental
measurement ($\varepsilon_{\infty}\thickapprox6.5$) \cite{lin}. The second
term stems from the rotations of MA ions. The dipole moment of the MA cation
is $p=7.64\times10^{-30}$C$\cdot$m and the number density of the cations is
$n_{d}\thickapprox$\ $4\times10^{27}$m$^{-3}$ [\onlinecite{fro}]. $\tau(T)$ is
the temperature dependent relaxation time of the MA ions \cite{pog} which is
about 0.2 -14 ps \cite{legu,pog,was}. The third term represents the
contribution of the displacements of Pb$^{2+}$ and I$^{-}$ ions, and the
factor $9/\beta(T-\theta)$ account for the static susceptibility
\cite{fey2,kit}. $\omega_{ip}$ is the frequency of ionic plasmon;
$\omega^{\prime}$\ and $\gamma^{\prime}$\ are the eigenfrequency and friction
coefficient of the Pb-I stretching mode \cite{ger}. $\beta$ is a constant
\cite{fey2} with a dimension of inverse temperature (K$^{-1}$). It turns out
that in MAPbI$_{3}$, $\theta\thicksim0$K is a small number \cite{sm} compared
to $T$. If the spontaneously polarized phase below the critical temperature is
ferroelectric, $\theta>0$. If the phase is anti-ferroelectric, $\theta<0$ [\onlinecite{kits}].

\begin{figure}
[ht]\centering{\includegraphics[width=0.34\textwidth]{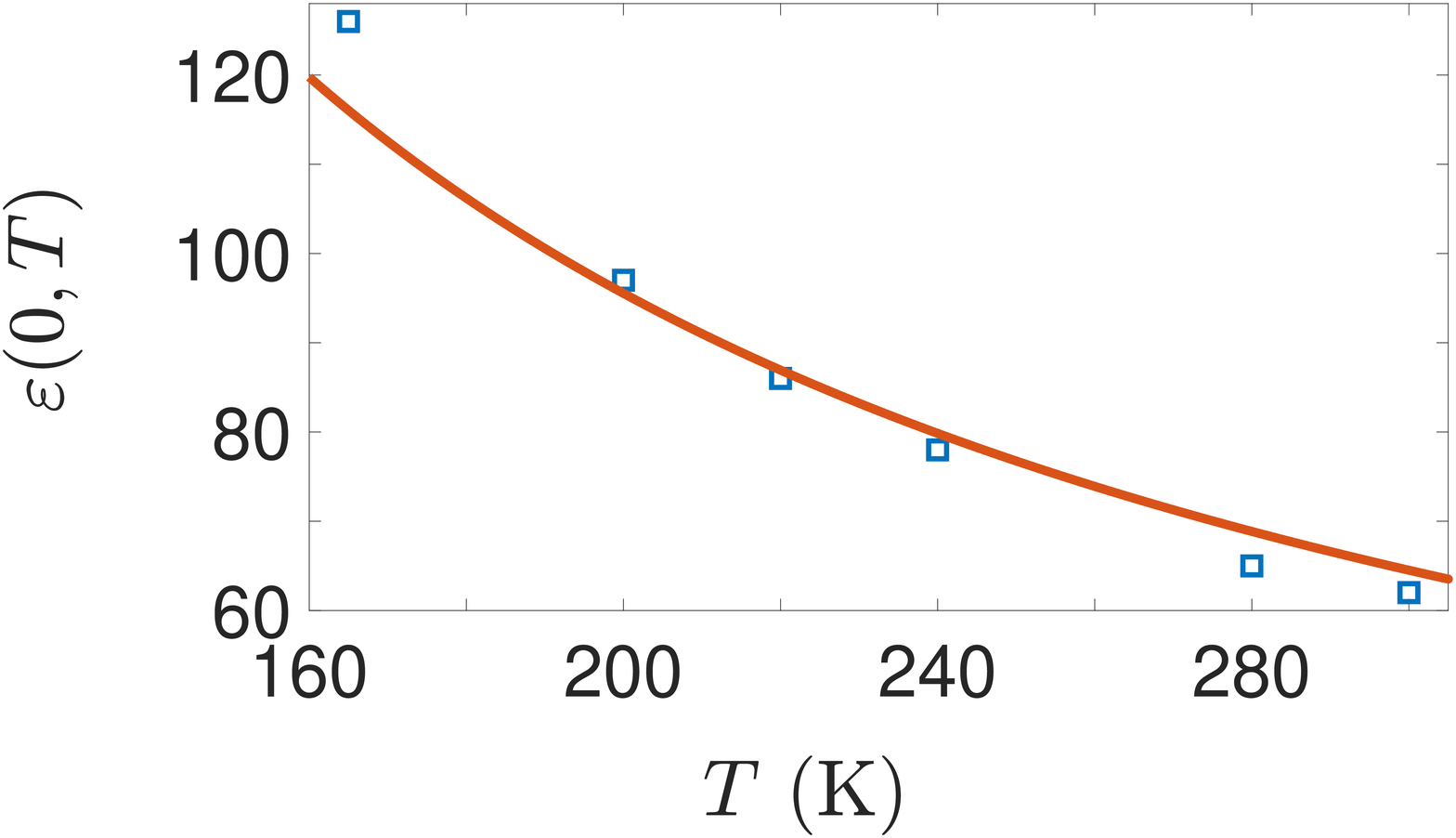}}
\caption{Static dielectric constant as function of temperature in MAPbI$_3$. The experimental data (squares) are taken from [\onlinecite{ono}], and the solid line is a fit from Eq. (\ref{die}).}
\label{DiT}
\end{figure}

If the frequency $\omega$ is so low ($\omega\ll7\times10^{10}$Rad/s) that the
product $\omega\tau(T)\ll1$, $[1-i\omega\tau(T)]^{-1}\thicksim1$. Hence the
second term reduces to $n_{d}p^{2}/(3k_{B}T\epsilon_{0})$. Specifically, at T
= 300K and $\omega=0$, the second terms becomes a constant (~2). Therefore,
the second term and the third term scale approximately as $1/T$, and
Eq.(\ref{die0}) becomes
\begin{equation}
\varepsilon(\omega,T)=\varepsilon_{\infty}+C(\omega)/T, \label{die}%
\end{equation}
where $C(\omega)$ is a materials constant, independent of temperature. This
result agree very well with the experimental data at $\omega/2\pi=1$KHz above
160K (cf. Fig. 3 of [\onlinecite{ono}]) as shown in Fig. \ref{DiT}. Note that
this $1/T$ dependence is analogous to Curie-Weiss law due to magnetic phase transitions.

If $\omega$ is ~$7\times10^{10}-10^{12}$Rad/s, $\omega\tau(T)\thicksim1$. In
this frequency range, the second term becomes a dominant contribution. As a
result, $\varepsilon(\omega,T)$ deviates from the $1/T$ behavior, as found
experimentally in the case of $\omega/2\pi= 90$GHz\ in [\onlinecite{pog}].

As will be shown later, the screened polaron-LA phonon interaction is
responsible for charge transport in the perovskites. The characteristic
acoustic phonon frequency is $\omega_{b}=c_{l}\pi/a$, where $a$\ is the
lattice constant and $c_{l}$\ is the speed of longitudinal sound wave. In
MAPbI$_{3}$, $\omega_{b}\thicksim10^{13}$Rad/s [\onlinecite{sm}]. For such a
high frequency, $\omega\tau(T)\gg1$, the second term can be ignored, and
Eq.(\ref{die0}) is reduced to Eq.(\ref{die}) again.

It is experimentally observed that below 150 -160 K, $\varepsilon(\omega,T)$
deviates from the $1/T$ behavior \cite{ono}, giving rise to contrasting
transport behaviors \cite{hut}. This deviation is caused by structural phase
transition from tetragonal to orthorhombic phase.

\section{charge transport and carrier mobility}

\subsection{Polaron-LA phonon vs. polaron-LO phonon interaction}

It is generally accepted that as quasiparticles, large polarons result from
the interaction between electrons (or holes) with optical phonons in ionic
perovskites \cite{zhu,bre,sen,pro,fro14}. However, it is often mistakenly
assumed that the same optical phonons must also be principally responsible for
scatting of the large polarons. This assumption would yield incorrect
temperature dependence of carrier mobility, which is the source of confusion
and debate in literature \cite{bre,zhu,wri,sen}. Here we demonstrate that much
of the electron-LO phonon interaction $h_{\text{e-LO}}$ is involved in the
formation of large polarons, thus the residual interaction $h_{\text{P-LO}}$
is substantially weaker than the interaction between the polaron and the LA
phonon, $h_{\text{P-Aph}}$ in MAPbI$_{3}$. Based on Born-Huang model of
electron-optical phonon interaction, we can show
\begin{equation}
h_{\text{e-LO}}/h_{\text{e-LA}}\thicksim\frac{\frac{3}{2}\hbar\omega
_{\text{LO}}(\frac{\hbar}{2m\omega_{\text{LO}}})^{1/4}[4\pi\alpha]^{1/2}%
}{\frac{1}{\varepsilon}\sqrt{\frac{\hbar}{2M_{j}c_{l}k_{b}}}2\frac{e^{2}%
}{\epsilon_{0}}n_{\text{cell}}^{1/2}}. \label{oa}%
\end{equation}
Here $n_{\text{cell}}$ is the number of primitive unit cells in a unit volume.
One can see that a softer lattice (smaller sound speed $c_{l}$), larger
primitive unit cell (smaller $k_{b}$ and $n_{\text{cell}}$) and smaller
$\omega_{\text{LO}}$ will increase the relative importance of $h_{\text{e-LA}%
}$. Since the electronic part of the polaron wave-function is similar to the
free electron wave-function, $h_{\text{P-LA}}\thickapprox h_{\text{e-LA}}$. On
the other hand, according to the Feynman model of large polarons,
\begin{equation}
h_{\text{P-LO}}\thicksim(\alpha/10)^{4}h_{\text{e-LO}}, \label{reo}%
\end{equation}
where $\alpha$\ is a dimensionless coupling constant \cite{feyp55}. Combining
Eq.(\ref{oa}) to Eq.(\ref{reo}), one has
\begin{equation}
\frac{h_{\text{P-LO}}}{h_{\text{P-LA}}}\thicksim(\alpha/10)^{4}\frac{\frac
{3}{2}\hbar\omega_{\text{LO}}(\frac{\hbar}{2m\omega_{\text{LO}}})^{1/4}%
[4\pi\alpha]^{1/2}}{\frac{1}{\varepsilon}\sqrt{\frac{\hbar}{2M_{j}c_{l}k_{b}}%
}2\frac{e^{2}}{\epsilon_{0}}n_{\text{cell}}^{1/2}}. \label{oac}%
\end{equation}
With the material parameters for MAPbI$_{3}$, we find $h_{\text{P-LO}%
}/h_{\text{P-LA}}\thicksim0.12$. This result is supported by the experiments
which reported \cite{wri,mene} $h_{\text{P-LO}}/h_{\text{P-LA}}\thicksim$ 0.1.
Therefore, the acoustic phonons are chiefly responsible for the scattering
while the optical phonons are responsible for the formation of the large
polarons in MAPbI$_{3}$. Recent experiments also suggest that the e-LO phonon
interaction is primarily responsible for the line-width of photoluminescence
(PL) spectrum of the perovskites \cite{wri}, which is consistent with the
preceding analysis. As mentioned earlier, since electrons and holes are
stabilized by the interaction with the optical phonons, they have to be
``activated" prior to recombination, by absorbing optical phonons. After the
annihilation, the optical phonons have to be emitted to restore the deformed
lattice. The energy of the absorbed and emitted phonons is responsible for the
PL line-width.

\subsection{Scattering mechanisms}

The Hamiltonian of the system can be written as
\begin{equation}
H=K_{\text{P}}+H_{\text{PP}}+H_{\text{P-def}}+H_{\text{P-LA}}+H_{\text{LA}},
\label{ha}%
\end{equation}
where $K_{\text{P}}$ denotes the sum of single polaron Hamiltonians, and
$H_{\text{PP}}$ is the Coulomb interaction between the polarons.
$H_{\text{P-def}}$ represents the interaction between the polarons and defects
whereas $H_{\text{P-LA}}$ is the interaction between the polarons and
longitudinal acoustic (LA) phonons; $H_{\text{LA}}$ is the Hamiltonian of LA
phonons. The interaction between the polarons and transverse phonons, and the
residual interaction between the polarons and LO phonons are small, and can be
neglected. Note that $H_{\text{PP}}$, $H_{\text{P-def}}$ and $H_{\text{P-LA}}$
represent dressed or effective interactions and are related to the
corresponding bare interactions via the dielectric function, e.g.,
$H_{\text{PP}}=H_{\text{PP}}^{\text{bare}}/\varepsilon(\omega,T)$.

We next apply the Boltzmann equation to elucidate the transport behavior of
large polarons in the perovskites. The key physical quantity of interest is
distribution function of the polarons, whose temporal rate change is given by
total collision frequency $\nu_{t}$, including scattering contributions of
polaron-polaron, polaron-defect and polaron-LA phonon. $\nu_{t}$ is related to
the charge carrier mobility $\mu$ by $\mu=e/m_{\text{P}}\nu_{t}$. We show that
the large polarons are stable against the three collision processes in
Supporting Information and to a good approximation, we can describe the
translational motion of the polarons by plane-waves. Thus the energy of the
polaron is given as $\varepsilon_{\mathbf{p}}=\mathbf{p}^{2}/2m_{\text{P}}$,
where $\mathbf{p}$ is the plane-wave momentum. We denote the non-equilibrium
distribution function of polarons in state $|\mathbf{p}\rangle$ as
$f_{\mathbf{p}}(t)$, and the distribution function of the LA phonons as
$N_{\mathbf{k}}(t)$ ($\mathbf{k}$ is the wave vector of the phonons) and their
corresponding equilibrium counterparts are given as $f_{0}$ and $N_{0}$.

The change rate of $f_{\mathbf{p}}(t)$ due to the polaron and LA phonon
collision is given by $\nu_{\text{P-LA}}=(\partial f_{\mathbf{p}}/\partial
t)_{\text{P-LA}}$ and is calculated in the following.
\begin{equation}
(\frac{\partial f_{\mathbf{p}}}{\partial t})_{\text{P-LA}}=-\sum_{\mathbf{k}%
}\frac{\partial N_{0}(\omega_{\mathbf{k}})}{\partial\hbar\omega_{\mathbf{k}}%
}[f_{0}(\mathbf{p}^{\prime})-f_{0}(\mathbf{p})] \label{ep1}%
\end{equation}%
\[
\{w(\mathbf{p}^{\prime},\mathbf{k};\mathbf{p})(\varphi_{\mathbf{p}^{\prime}%
}-\varphi_{\mathbf{p}}+\chi_{\mathbf{k}})\delta(\varepsilon_{\mathbf{p}%
}-\varepsilon_{\mathbf{p}^{\prime}}-\hbar\omega_{\mathbf{k}})
\]%
\[
-w(\mathbf{p}^{\prime};\mathbf{p},\mathbf{k})(\varphi_{\mathbf{p}^{\prime}%
}-\varphi_{\mathbf{p}}-\chi_{\mathbf{k}})\delta(\varepsilon_{\mathbf{p}%
}-\varepsilon_{\mathbf{p}^{\prime}}+\hbar\omega_{\mathbf{k}})\}.
\]
Here $\varphi$ and $\chi$ describe the deviations of $f_{\mathbf{p}}$ and
$N_{\mathbf{k}}$ from their equilibrium values: $f_{\mathbf{p}}-f_{0}%
(\varepsilon)=-\varphi\partial f_{0}(\varepsilon)/\partial\varepsilon$, and
$N_{\mathbf{k}}-N_{0}(\mathbf{k})=-\chi\partial N_{0}(\omega_{\mathbf{k}%
})/\partial\hbar\omega_{\mathbf{k}}$. $w(\mathbf{p}^{\prime},\mathbf{k}%
;\mathbf{p})$ is the probability amplitude defined as $w(\mathbf{p}^{\prime
},\mathbf{k};\mathbf{p})(N_{\mathbf{k}}+1)=2\pi|\langle\mathbf{p}^{\prime
},\mathbf{k}|H_{\text{P-LA}}($emission$)|\mathbf{p}\rangle|^{2}/\hbar$,
$w(\mathbf{p}^{\prime};\mathbf{p},\mathbf{k})$ is defined by $w(\mathbf{p}%
^{\prime};\mathbf{p},\mathbf{k})N_{\mathbf{k}}=2\pi|\langle\mathbf{p}^{\prime
}|H_{\text{P-LA}}($absorption$)|\mathbf{p},\mathbf{k}\rangle|^{2}/\hbar$
[\onlinecite{v10}]. Similar rate equations can be obtained for $\nu
_{\text{P-def}}$ and $\nu_{\text{PP}}$ and their expressions are given in
Supporting Information.

The characteristic frequency of the LA phonons is $\omega_{b}=c_{s}k_{b}$,
where $c_{s}$ is the average sound speed in the longitudinal direction.
$k_{b}=\pi/a$ is the wave-vector at the Brillouin zone boundary \cite{cal}.
Because the elastic constants of the perovskites are relatively small, $c_{s}$
and $\omega_{b}$ are also small. In the tetragonal phase \cite{he,qia} of
MAPbI$_{3}$, $c_{s}\approx$ 2147 m/s, and $\omega_{b}\sim$ 82 K. In the
pseudo-cubic phase \cite{he,qia}, $c_{s}\approx2824$ m/s, $\omega_{b}%
\thicksim$ 107 K. Thus at room temperature, $k_{B}T\gtrsim\hbar\omega_{b}$ and
the LA phonons are fully excited \cite{hut}. These fully excited LA phonons
increase the P-LA scattering probability and are principally responsible for
polaron scattering. In addition, the phonon distribution function
$N_{0}(\omega_{\mathbf{k}})$ in Eq. (\ref{ep1}) can be reduced to
$N_{0}(\omega_{\mathbf{k}})\thickapprox k_{B}T/\hbar\omega_{\mathbf{k}}$.

We can now derive an analytical expression for the change rates $\partial
f_{\mathbf{p}}/\partial t$ induced by the three collision processes
$H_{\text{PP}}$, $H_{\text{P-def}}$ and $H_{\text{P-LA}}$. More specifically,
change rate due to the polaron-polaron scattering is given by \cite{pei,sm}
$\nu_{\text{PP}}=(\partial f_{\mathbf{p}}/\partial t)_{\text{PP}}$:
\begin{equation}
\nu_{\text{PP}}\thicksim\left[  \frac{T}{300\varepsilon_{s1}%
+(T-300)\varepsilon_{\infty}}\right]  ^{2}n \label{ee}%
\end{equation}%
\[
\frac{4\pi^{3/2}\hbar^{3}e^{-3/2}}{(2m_{\text{P}}k_{B}T)^{3/2}}\frac{1}{\hbar
}\frac{d^{4}}{a^{6}}(\frac{e^{2}}{\epsilon_{0}})^{2}\frac{(kT)^{2}}{D^{3}},
\]
where the dielectric function $\varepsilon_{s1}=\varepsilon(\omega_{b},300)$;
$D\thicksim$ 3 eV is the conduction band width \cite{bri,uma,pro} of
MAPbI$_{3}$ and $d=2R_{\text{P}}$ is the diameter of the polaron. The change
rate due to the polaron-defect scattering is \cite{v10,cal,sm} given by
$\nu_{\text{P-def}}=(\partial f_{\mathbf{p}}/\partial t)_{\text{P-def}}$ :
\begin{equation}
\nu_{\text{P-def}}\thicksim\left[  \frac{T}{300\varepsilon_{s1}%
+(T-300)\varepsilon_{\infty}}\right]  ^{2}C\frac{2\pi}{\hbar} \label{cdf1}%
\end{equation}%
\[
\left(  \frac{e^{2}\Delta z}{\epsilon_{0}}\right)  ^{2}\frac{1}{D^{2}a^{3}%
}\frac{\hbar^{4}}{(2m_{\text{P}}k_{B}T)^{2}}k_{B}T,
\]
where $C$ is the number of defects per cubic meter and $\Delta z$ is the
effective charge of the defect. The change rate due to polaron and LA phonon
scattering is given by $\nu_{\text{P-LA}}=(\partial f_{\mathbf{p}}/\partial
t)_{\text{P-LA}}$:
\[
\nu_{\text{P-LA}}\thicksim\left[  \frac{T}{300\varepsilon_{s1}%
+(T-300)\varepsilon_{\infty}}\right]  ^{2}\frac{\pi}{M\omega_{b}k_{b}^{2}%
a^{3}}%
\]%
\begin{equation}
\frac{4\pi}{3}k_{b}^{3}(\frac{ze^{2}}{\epsilon_{0}})^{2}(\frac{k_{B}T}%
{\hbar\omega_{b}})^{2}n\frac{4\pi^{3/2}\hbar^{3}e^{-3/2}}{(2m_{\text{P}%
})^{3/2}(k_{B}T)^{5/2}}, \label{cip}%
\end{equation}
where $z$ is the weighted nuclear charges of the ions and $M$ is the reduced
mass of Pb and I ions.

It is known that dominant defects in halide perovskites are not particularly
harmful to charge transport because they do not create detrimental deep levels
within the band gap \cite{RevYin,yinAPL,miller}. Therefore, in our model, only
shallow defects are considered, which could induce lattice deformation and
charge states at the defect center. Because polaron scattering due to the
former is much smaller than the latter, we can approximate $H_{\text{P-def}}$
by Coulomb interaction between the point charges at the defect center and the polarons.

To compare the relative importance of the scattering processes, we evaluate
the three terms by taking I$^{-}$ vacancies as an example of defects in
MAPbI$_{3}$. We assume a moderate defect concentration at $C=4.0\times10^{20}%
$cm$^{-3}$ and $\Delta z=1.22$. The consideration of other point defects will
only change $\Delta z$ by a small amount ($\Delta$z = 1 - 3). The three
contributions as a function of temperature are plotted in Fig. \ref{EPC}. We
find that at room temperature $\nu_{\text{P-Aph}}\gg\nu_{\text{P-def}}\gg
\nu_{\text{PP}}$. Therefore, the polaron-LA phonon scattering dominates charge
transport in MAPbI$_{3}$, and $\mu$ would appear insensitive to the defects.

\begin{figure}
[ht]\centering
\subfigure[]{\includegraphics[width=0.23\textwidth]{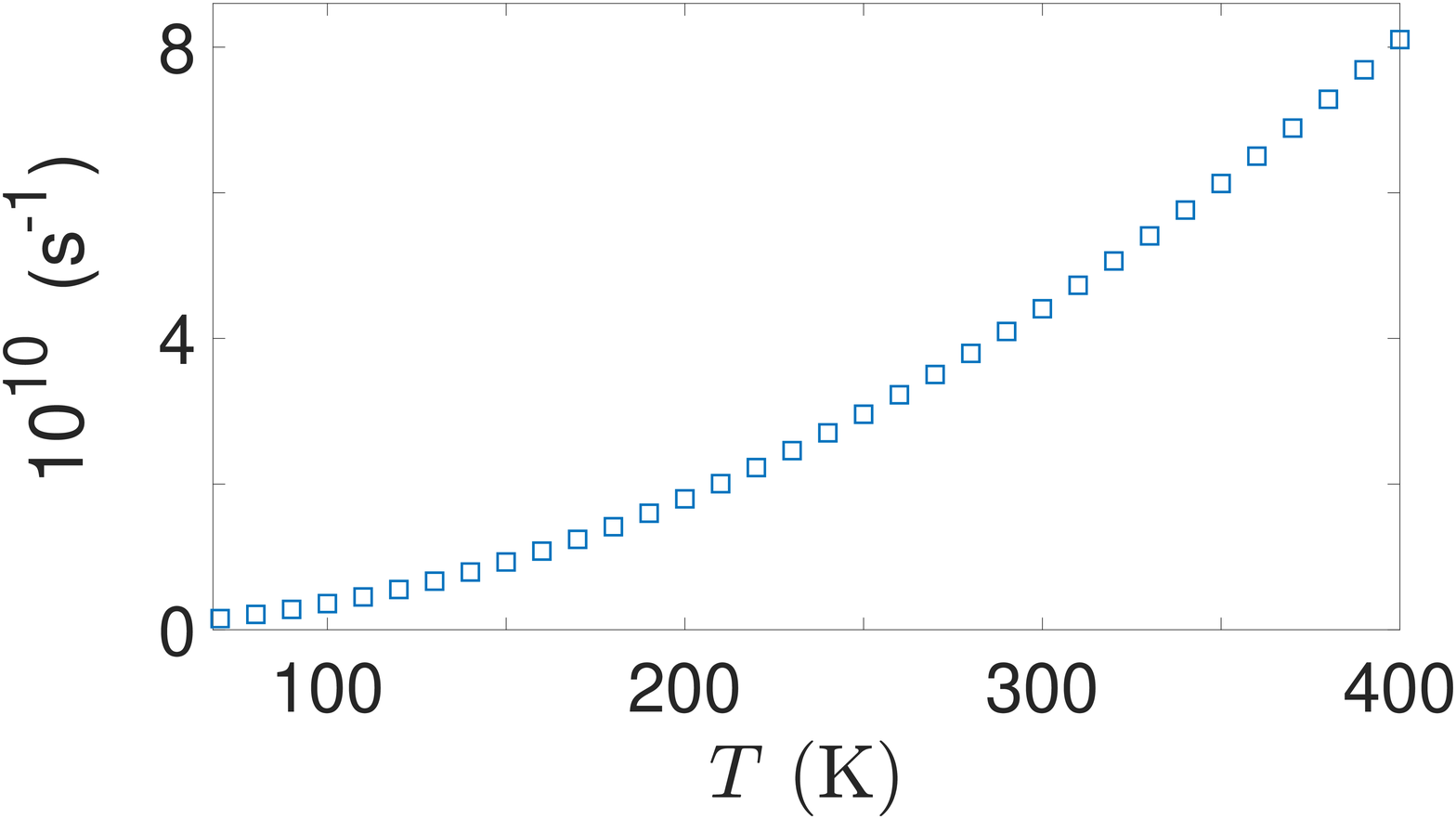}\label{eeP}}
\subfigure[]{\includegraphics[width=0.23\textwidth]{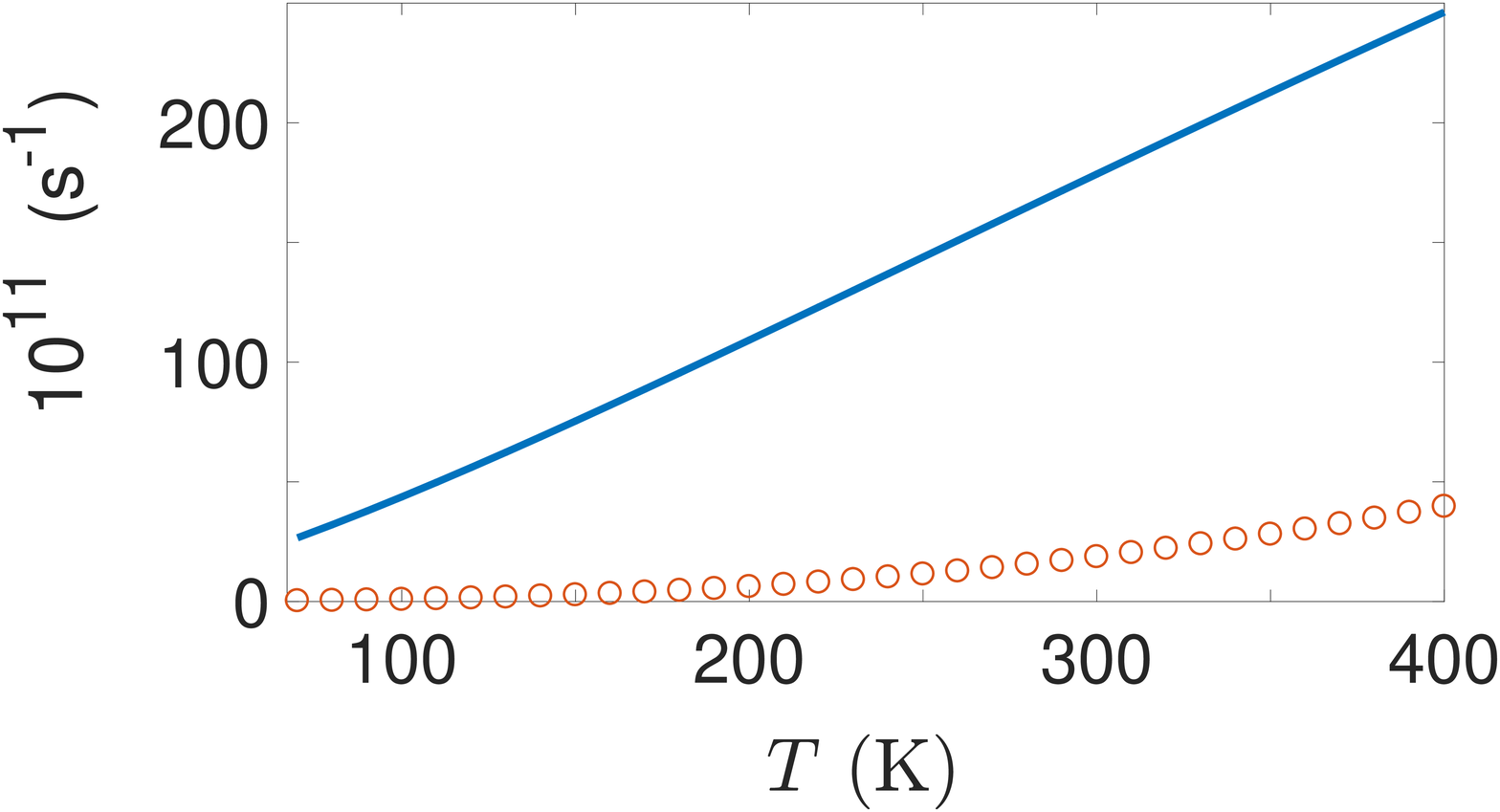}\label{vaca}}
\caption{(a) The polaron-polaron collision frequency as a function of $T$ determined by Eq.(\ref{ee}); (b) The polaron-I$^{-}$ vacancy collision frequency (circles) and the polaron-LA phonon collision frequency (solid line) as functions of T determined by Eq. (\ref{cdf1}) and (\ref{cip}).
$\varepsilon_{0}=70$ and $\varepsilon_{\infty}=6.5$ [\onlinecite{lin}] are used in the plot.}
\label{EPC}
\end{figure}

\subsection{Concentration dependence of mobility}

\begin{figure}
[ht]\centering
\subfigure[]{\includegraphics[width=0.23\textwidth]{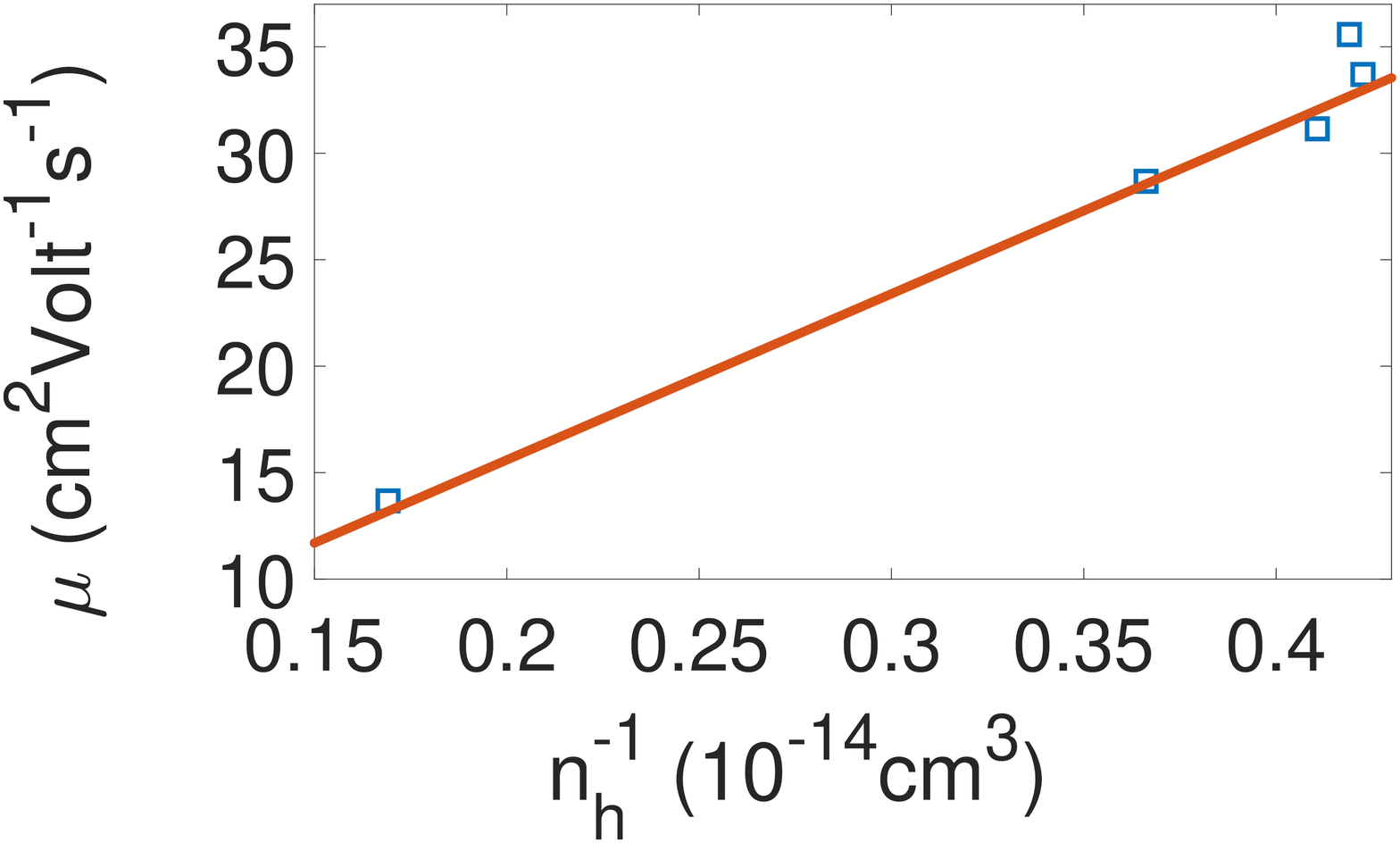}\label{hmc}}
\subfigure[]{\includegraphics[width=0.23\textwidth]{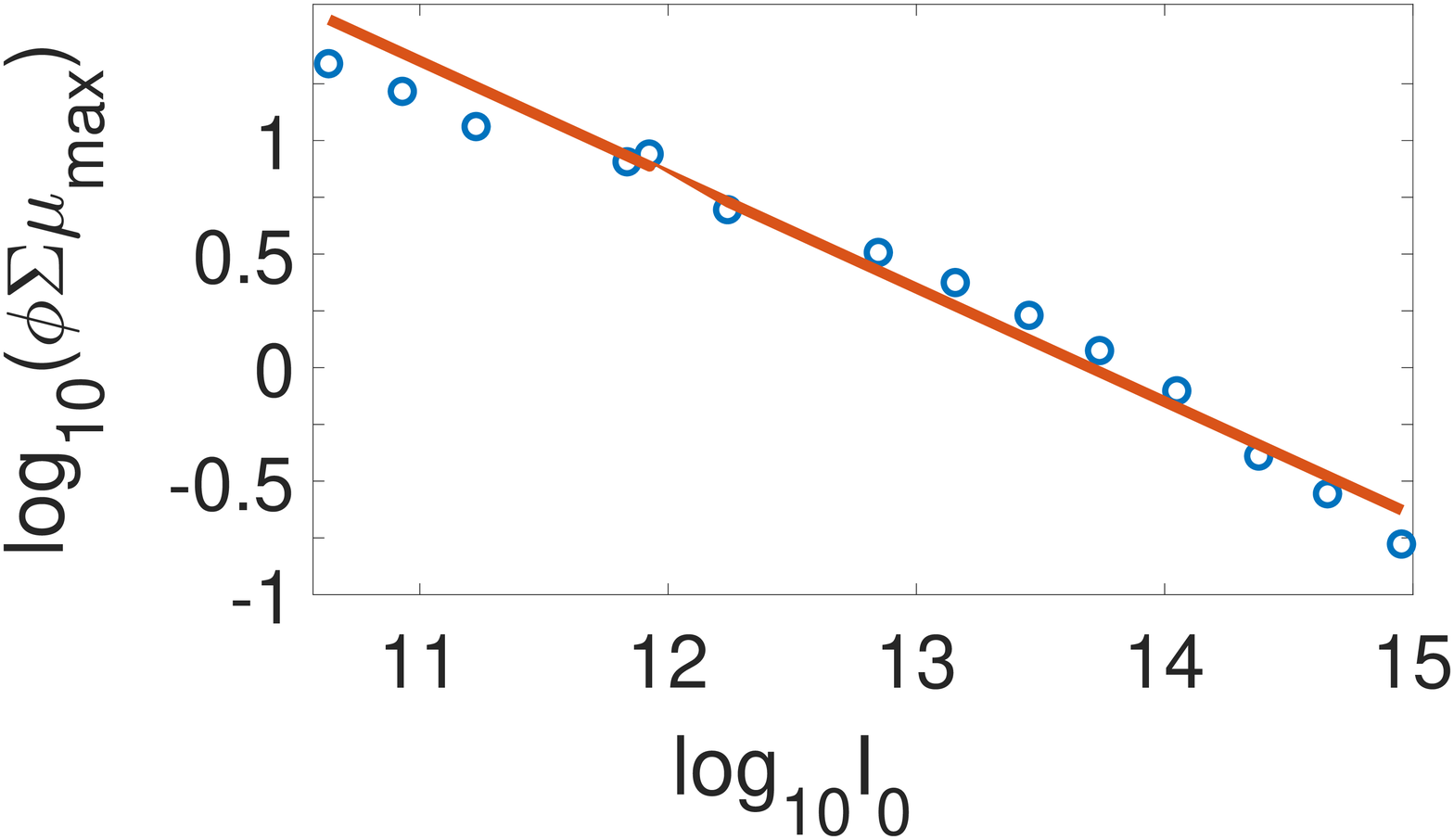}\label{flux}}
\caption{(a) Hole mobility $\mu$ vs. $n^{-1}_{h}$ for p-doped MAPbI$_3$. The experimental data (squares) are taken from [\onlinecite{bi}]; the solid line is a fit of Eq. (\ref{epf}). (b) The logarithm of the effective carrier mobility, log$_{10}\phi\Sigma\mu$ is plotted as a function of the logarithm of incident photon flux, log$_{10}I_{0}$.
The experimental data (circles) are taken from [\onlinecite{oga}], and the solid line is a fit of Eq. (\ref{i0}).}
\label{mc}
\end{figure}

If we ignore $\nu_{\text{P-def}}$ and $\nu_{\text{PP}}$, we arrive at the key
result of the model:
\begin{equation}
\mu\varpropto n^{-1}m_{\text{P}}^{1/2}T^{-3/2}. \label{epf}%
\end{equation}
First, we find that the mobility is inversely proportional to the carrier
concentration $n$, and this finding is consistent to the experimental
measurements \cite{bi} in p-doped MAPbI$_{3}$. In Fig. \ref{hmc}, we compare
the experimental hole mobility $\mu_{h}$ (squares) with the theoretical values
(solid line) as a function of $n^{-1}$ where a good agreement is found.

Let $\gamma$ be the electron-hole recombination coefficient, $\kappa$ the
generation probability per impinging photon and $G$ the volume density of
photons in the sample, we can express $n=(\gamma^{-1}\kappa G)^{1/2}$ by
assuming $n$ is much larger than the trap center concentration. Here,
$G=I_{0}/l_{\text{abs}}$, $I_{0}$ is the incident photon flux and
$l_{\text{abs}}$ is the absorption length \cite{che}. Substitute the
expression of $n$ into Eq. (\ref{epf}), one obtains:
\begin{equation}
\mu\varpropto(l_{\text{abs}}\gamma)^{1/2}(\kappa I_{0})^{-1/2}T^{-3/2}.
\label{i0}%
\end{equation}
The circles in Fig. \ref{flux} are experimental data \cite{oga} for effective
mobility $\phi\mu$ vs. incident flux $I_{0}$, and the solid line is a fit
based on Eq. (\ref{i0}). Here we have to adjust the intercept due to a lack of
experimental values of $\kappa$, $l_{\text{abs}}$ and $\gamma$ in
[\onlinecite{oga}], nevertheless the agreement in the slope between the theory
and the experiment is very good.

\subsection{Temperature dependence of mobility}

\begin{figure}
[ht]\centering
\subfigure[]{\includegraphics[width=0.225\textwidth]{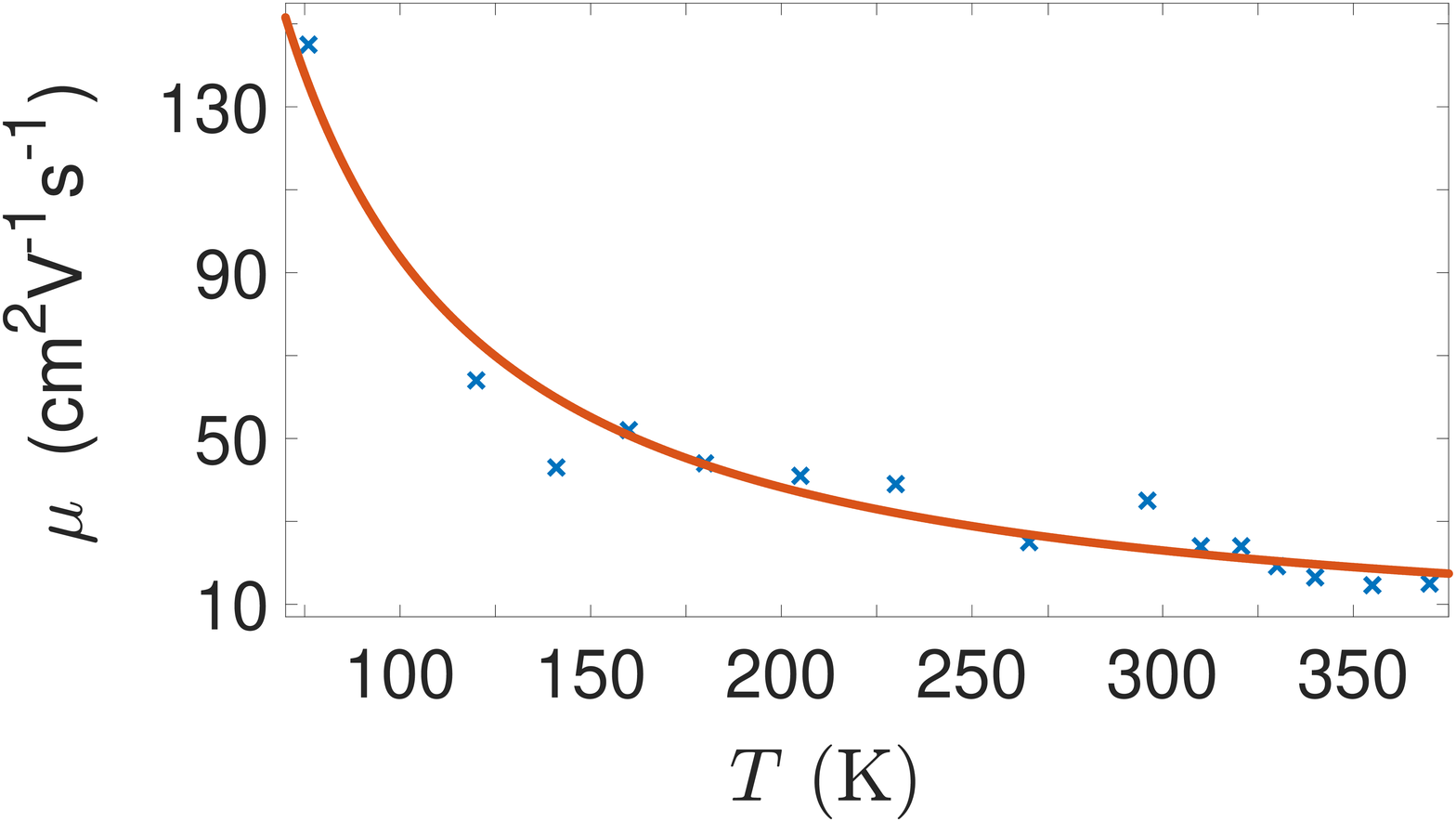}
\label{milExa245}}
\subfigure[]{\includegraphics[width=0.225\textwidth]{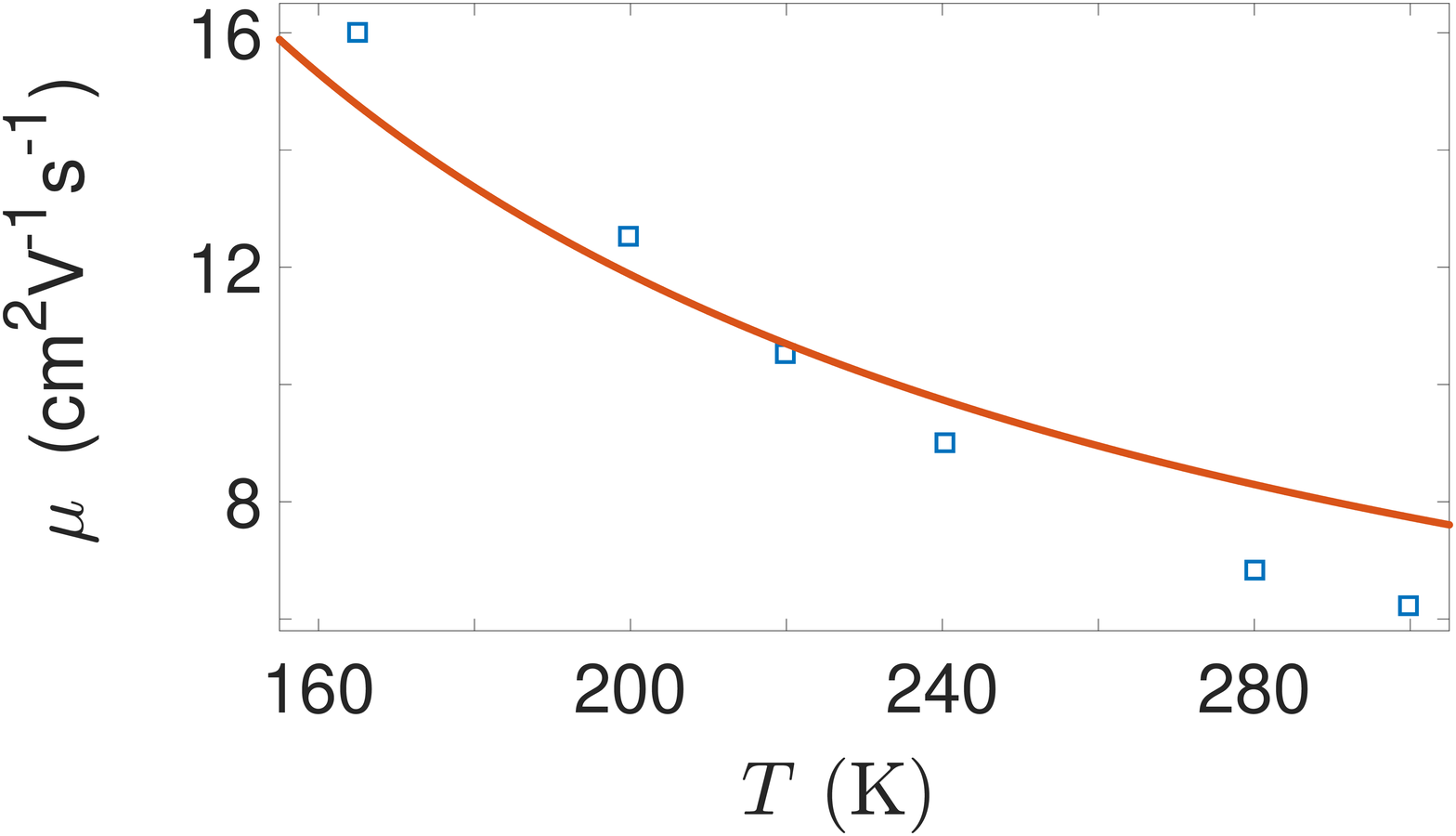}
\label{SavExa245}}
\caption{The carrier mobility $\mu$ as a function of temperature. The solid curves are obtained from Eq. (\ref{cip}) with fitting a parameter of $n$.
(a) The experimental data (crosses) is from [\onlinecite{mil}], and $n=2.3\times 10^{17}$cm$^{-3}$. (b) The experimental
data (squares) is from \onlinecite{sav}, and $n=8.3\times 10^{17}$cm$^{-3}$.}\label{mob245}%

\end{figure}

Finally, we compare the theoretical prediction with experimental data on
carrier mobility as a function of temperature making use of Eq. (\ref{cip})
and $\mu(T)=e/m_{\text{P}}\nu_{\text{P-LA}}$. Because the values of
$\varepsilon_{0}$, $\varepsilon_{\infty}$ and $n$ are not available in the
experiments \cite{mil,sav}, we have to use $n$ as a fitting parameter in the
comparison. By taking $\varepsilon_{\infty}$ = 4.5 and $\varepsilon_{0}$ =
24.5 from first-principles calculations \cite{fro14,bri}, we can fit the
theoretical mobility to the experimental data in Fig. \ref{mob245}. For the
first experiment \cite{mil}, $n=2.3\times10^{17}$cm$^{-1}$ was used in the
fitting while for the second experiment \cite{sav}, $n=8.3\times10^{17}%
$cm$^{-1}$ was used in the fitting. Both values of $n$ are reasonable
\cite{che} and for both cases, satisfactory agreements to the experimental
data are obtained.

\begin{figure}
[ht]\centering{\includegraphics[width=0.34\textwidth]{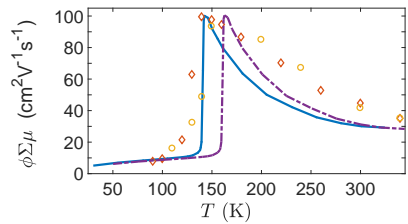}}
\caption{Product of generation yield $\phi$ and mobility $\Sigma\mu$ as function of temperature. Circles (diamonds) are measured during
the heating (cooling) process [\onlinecite{hut}]. The dash line is calculated from the experimental dielectric function [\onlinecite{ono}] and the solid line is obtained by shifting the
the tetragonal-orthorhombic transition temperature from [\onlinecite{ono}] to [\onlinecite{hut}].}
\label{HutOn}
\end{figure}

In a recent experiment by Hutter et al., the temperature dependence of carrier
mobility in MAPbI$_{3}$ was shown to exhibit two regimes of contrasting
behaviors \cite{hut}. Above 150 K, carrier mobility $\mu(T)\varpropto
T^{-3/2}$ while below 150 K, the mobility drops precipitately, decreasing with
decreased temperature. Using the experimental dielectric function
$\varepsilon(\omega,T)$ for $\omega/2\pi=1$KHz as obtained in \cite{ono}, our
analytical expression in Eq. (\ref{cip}) can reproduce the experimental data
of Hutter reasonably well in both regimes, as shown in Fig.\ref{HutOn}. In the
tetragonal phase ($T>$ 150 K), the mobility behaves as $\mu(T)\varpropto
T^{-3/2}$, while in the orthorhombic phase ($T<$ 150K), the mobility decreases
with decreasing temperature with a sharp drop around 150 K. We further
speculate that the reason that the earlier experiments observed only the
regime of $\mu(T)\varpropto T^{-3/2}$ is due to the presence \cite{kong,hut}
of the tetragonal phase at $T<$ 150 K.

\section{Summary}

In conclusion, we proposed a theoretical model that can elucidate key
experimental observations on charge transport in hybrid perovskite materials.
Essential to the model is improved understanding crucial to charge transport,
including that the acoustic phonons as opposed to the optical phonons are
responsible for the scattering of large polarons, the acoustic phonons are
fully excited due to the \textquotedblleft softness\textquotedblright of the
perovskites, and the temperature dependent dielectric function is the key
contributor to the temperature dependence of the mobility. Analytic
expressions were given for various contributions to the carrier mobility and
compared to the experimental measurements with good agreements. By directly
relating the carrier mobility to material parameters, the present work may
provide guidance for materials design and form a starting point for more
rigorous first-principles predictions of transport properties.

The work at California State University Northridge was supported by NSF-PREM
program via grant DMR-1205734. Discussion with Guangjun Nan is acknowledged.
The authors wish to thank anonymous referees for their stimulating comments.
\newline$^{\dagger}$haiqing0@csrc.ac.cn, $^{\ast}$ganglu@csun.edu

\begin{widetext}
\textbf{Supplemental Material for}\newline
\begin{center}
{\large\textbf{Charge Transport in Hybrid Halide Perovskites}}\newline
Mingliang Zhang$^{1,2}$, Xu Zhang$^{2}$, Ling-Yi Huang$^{2}$,
Hai-Qing Lin$^{1}$ and Gang Lu$^{2}$\newline
$^{1}$Beijing Computational Science Research Center, Beijing 100193, China
\newline
$^{2}$Department of Physics and Astronomy, California State University
Northridge, Northridge, CA 91330, USA
\end{center}
\end{widetext}

\subsection{dielectric screening}

In MAPbI$_{3}$, there are four factors contributing to dielectric function
$\varepsilon(\omega,T)$: (i) bound electrons; (ii) the displacements of
Pb$^{2+}$ and I$^{-}$ ions of the lattice frame; (iii) the rotation of MA
dipoles and (iv) free electrons. The contribution (iv) of `free' carriers is
negligible. We will see that the displacements of ions is the most important.

\subsubsection{MA\ dipoles}

Let us show that the MA dipoles cannot form a spontaneous polarized phase.
Both experiments \cite{legu} and simulations \cite{bri13, mos14, qua15} show
that the rotational barrier for the changing direction of a MA dipole is
larger than 13.5meV (157K): below 157K, the MA dipoles are locked in various
orientations, i.e. a spontaneous polarization cannot be implemented by the MA
dipoles below 157K. In addition, the attraction energy between two dipoles is
the strongest if they take the same direction and are parallel to the
connection line of their centers \cite{ger}:%
\begin{equation}
U=-\frac{2}{4\pi\epsilon_{0}\varepsilon_{\infty}}\frac{p_{1}p_{2}}%
{|\mathbf{x}_{1}-\mathbf{x}_{2}|^{3}},\label{sat}%
\end{equation}
where $\mathbf{x}_{1}$ and $\mathbf{x}_{2}$ are the position vectors of two
dipoles $\mathbf{p}_{1}$ and $\mathbf{p}_{2}$. In MAPbI$_{3}$, the nearest
distance between two MA dipoles is $|\mathbf{x}_{1}-\mathbf{x}_{2}|=6.3$\AA ,
$\varepsilon_{\infty}=6.5$ [\onlinecite{lin}], the dipole moment of MA$^{+}$
is \cite{fro} $p=7.64\times10^{-30}$C$\cdot$m, then $U=4$meV=47K$\ll
$25meV=300K. At $T>47$K, the MA dipoles cannot align in the same direction to
form a ferroelectric. Therefore, only the wobbling and rotating of the MA
dipoles contribute to the dielectric polarization.

\subsubsection{Spontaneous polarization at low temperature}

\label{disspo}

We give a simple reasoning to support that the existence of a spontaneously
polarized phase at low temperature is caused by the displacements of Pb$^{2+}
$ and I$^{-}$ ions. The perovskite crystal can be viewed as composed of three
types of chains: (1) ---I$^{-}$---Pb$^{2+}$---I$^{-}$---; (2) ---I$^{-}
$---I$^{-}$---; and (3) ---MA---MA---. Denote the lattice constant of
MAPbI$_{3}$ as $a$, the distance between two MAs is $a$, the distance between
two I$^{-}$ ions is $a$, the distance between I$^{-}$ and Pb$^{2+}$ is $a/2$.
If one applies an electric field along the direction of these chains, ions
will move accordingly. Since the field produced by a dipole is proportional to
the inverse cube of distance, in a processes of spontaneous polarization, we
may ignore the---I$^{-}$---I$^{-}$--- chains and the ---MA---MA--- chains. For
an ion in a given chain, the field at the position of that ion produced by
other ---I$^{-}$---Pb$^{2+}$---I$^{-}$--- chains is much weaker than the field
produced by the ions in the same chain \cite{fey2}. Thus we only need to
consider one ---I$^{-}$---Pb$^{2+}$---I$^{-}$--- chain. In a perovskite
structure, the critical polarizability $\alpha_{c}$ for the existence of a
spontaneously polarized (ferroelectric or anti-ferroelectric) phase is
\cite{fey2}: $\alpha_{c}=(a/2)^{3}/0.383$. In MAPbI$_{3}$, $a=6.3$\AA , then
$\alpha_{c}=8.16\times10^{-29}$m$^{3}$.

The electronic polarization of I$^{-}$ and Pb$^{2+}$ ions, and the orientation
polarization of the MA dipoles are not enough to produce a spontaneous
polarization. The polarizabilities of I$^{-}$ and Pb$^{2+}$ are $\alpha
_{\text{I}^{-}}=6.43\times10^{-30}$m$^{3}$, $\alpha_{\text{Pb}^{2+}}%
=4.9\times10^{-30}$m$^{3}$\cite{tes}, which are far from enough to produce a
spontaneous polarization. If the MA dipole can rotate freely, the average
dipole $\langle p\rangle_{T}$ at temperature $T$ is \cite{ger}%
\begin{equation}
\langle p\rangle_{T}\thickapprox\frac{p^{2}}{3k_{B}T\epsilon_{0}}\epsilon
_{0}E,\label{avd}%
\end{equation}
where $E$ is the strength of electric field. Then the rotational
polarizability $\alpha_{\text{MA}}$ of MA dipoles is%
\begin{equation}
\alpha_{\text{MA}}=\frac{p^{2}}{3k_{B}T\epsilon_{0}}.\label{pma}%
\end{equation}
One can see that when $T\lesssim44$K, $\alpha_{\text{MA}}$ could reach
$\alpha_{c}$. However, we overestimated $\alpha_{\text{MA}}$. In the
fabrication process, the orientations of MA dipoles are random. The energy
barrier for the reorientation of a MA dipole is at least 13.5 meV (157.2K)
\cite{fro,leg}. When the temperature is lowered to $T\lesssim44$K, the MA
dipoles are already locked into various orientations by the energy barrier. A
spontaneous polarization cannot be produced from the reorientation of the MA dipoles.

Let us consider the induced dipole caused by the displacements of Pb$^{2+}$
and I$^{-}$ ions. To make Pb$^{2+}$ sit in the octahedral hole of I$^{-}$, one
requires that $0.414R<r<0.732R$ [\onlinecite{oct}], where $r=1.33$\AA \ is the
radius of Pb$^{2+}$, $R=2.06$\AA \ is the radius of I$^{-}$ [\onlinecite{ir}].
One can see $r/R=0.65$, Pb$^{2+}$ is not enclosed by the I$^{-}$ ions too
tightly. Denote the spring constant of Pb$^{2+}$---I$^{-}$ as $k$, the charge
of I$^{-}$ as $q_{\text{I}^{-}}$, the charge of Pb$^{2+}$ as $q_{\text{Pb}%
^{2+}}$. The stretch frequency of Pb$^{2+}$---I$^{-}$ is $\widetilde{\nu}%
=$106.9 cm$^{-1}$ [\onlinecite{perez}]. Then $k=m_{r}\omega^{2}\thicksim
53$Nm$^{-1}$, where $m_{r}$ is the reduced mass of Pb$^{2+}$ and I$^{-}$. In
an electric field $\mathbf{E}$, the induced dipole $\mathbf{p}_{in}%
=q_{\text{Pb}^{2+}}d_{2}+q_{\text{I}^{-}}d_{1}$ of ---Pb$^{2+}$---I$^{-}$---
\ is%
\begin{equation}
\mathbf{p}_{in}=\frac{q_{\text{Pb}^{2+}}^{2}+q_{\text{I}^{-}}^{2}}%
{k\epsilon_{0}}\epsilon_{0}E,\label{ind}%
\end{equation}
where $d_{1}$ and $d_{2}$ are the induced displacements of I$^{-}$ ion and
Pb$^{2+}$ ion. The polarizability $\alpha_{\text{dis}}$ due to the
displacements of Pb$^{2+}$ and I$^{-}$ ions is%
\begin{equation}
\alpha_{\text{dis}}=\frac{q_{\text{I}^{-}}^{2}+q_{\text{Pb}^{2+}}^{2}%
}{\epsilon_{0}k}\thickapprox2.73\times10^{-28}\text{m}^{3}>\alpha
_{c}.\label{pis}%
\end{equation}

Above estimation only considered the induced displacements of Pb$^{2+}$ and
I$^{-}$ ions without the perturbation of thermal vibrations. Considering
$\alpha_{\text{dis}}$ is only three times larger than $\alpha_{c}$, thermal
vibrations could significantly reduce $d_{1}$ and $d_{2}$, i.e $\alpha
_{\text{dis}}$. It seems reasonable to assume that at a low enough
temperature, MAPbI$_{3}$ is in a spontaneously polarized phase (either in a
ferroelectric phase or in an anti-ferroelectric phase). To decide if a
ferroelectric or an anti-ferroelectric is more favorable, one needs a more
refined calculation to determine the direction of the internal field on the
---I$^{-}$---I$^{-}$--- chain is whether parallel or antiparallel to the field
on the ---I$^{-}$---Pb$^{2+}$---I$^{-}$--- chain.

\subsubsection{Neglect of the susceptibility from free carriers}

\label{fc}

The screening caused by the `free' carriers may affect the properties of
ABX$_{3}$ in two aspects: (a) screening the the electric field $E_{\text{pol}%
}$ which causes spontaneous polarization; (b) screening three bare
interactions: polaron-polaron (PP) interaction, polaron-defect interaction
(P-def), and polaron-acoustic phonon interaction (P-LA) \cite{pine}.

Let us first discuss effect (a). The maximal screening is reached if the
positive and negative charges are concentrated in two opposite surfaces. The
observed size $l$ of a ferroelectric domain is $l\lesssim10^{-6}$m
\cite{kut,fan,kim}, then the surface charge density $\sigma$ is $\sigma
\thicksim nel$. The electric field $E_{\text{res}}$ produced by the free
carriers is $E_{\text{res}}\thicksim nel/\epsilon_{0}$. The induced dipole
$p_{\text{res}}$ by the field $E_{\text{res}}$ of free carriers is
$p_{\text{res}}\thicksim\alpha_{\text{cri}}nle$. If $p_{\text{res}}$ is
comparable to the dipole $p($MA$)$ of MA, then the spontaneous polarization is
modified. For $l=10^{-6}$m, the upper limit concentration is $n^{\text{upper}%
}=p($MA$)/\alpha_{\text{cri}}le\thicksim5.85\times10^{17}$cm$^{-3}$.
Considering we used the largest $l$, the actual $n^{\text{upper}}$ may be
higher than above value. Then, for $n_{e}<10^{18}$cm$^{-3}$, the changes in
elastic constants and $E_{\text{pol}}$ are negligible. ABX$_{3}$ is still in a
super paraelectric phase.

Secondly, let us discuss effect (b). For $2\times10^{17}\lesssim n_{e}%
\lesssim10^{18}$cm$^{-3}$ and $T\lesssim$350K the Debye-Huckel screening
length $r_{\text{DH}}$ is $r_{\text{DH}}=\sqrt{\epsilon_{0}k_{B}T/ne^{2}%
}\lesssim28.9$\AA . The diameter of an EP is $2R_{\text{EP}}$ $\thicksim
56.7$\AA \ $\gtrsim$ $r_{DH}$. Debye-Huckel type electronic screening does not
happen; for $n_{e}<2\times10^{17}$cm$^{-3}$ and $T>350$K, $r_{\text{DH}}%
^{-1}=(ne^{2}/\epsilon_{0}k_{B}T)^{1/2}\ll k_{b}$, the Debye-Huckel screening
plays little role except for the phonons with very low frequencies \cite{zim}.
But if $n>10^{18}$cm$^{-3}$, the carriers are electrons (holes), the
Thomas-Fermi screening caused by the `free' electrons.is significant.

In summary, if $I_{0}$ is not too big, i.e. $n_{e}<10^{18}$cm$^{-3}$, the
screening mainly comes from the bound electrons and the displacements of ions.

\subsection{Screened interactions}

The effective interaction $H_{\text{e-LA}}$ of electron-LA phonon relates to
the bare interaction by \cite{pine}:
\begin{equation}
H_{\text{e-LA}}=H_{\text{e-LA}}^{\text{bare}}/\varepsilon(\omega,T).\label{ef}%
\end{equation}
To get a manageable expression for mobility $\mu$, we use%

\begin{equation}
\varepsilon(\omega_{b},T)=\varepsilon_{\infty}+\frac{C(\omega_{b})}%
{T},\label{x0}%
\end{equation}
at the characteristic frequency $\omega_{b}=c_{s}k_{b}\thicksim1.41\times
10^{13}$rad/s of P-Aph interaction to all frequency $\omega$. Denote
$\varepsilon_{s1}=\varepsilon(\omega_{b},T_{1})$ at a specific temperature
$T_{1}$, then for temperature $T$, one has%
\begin{equation}
\frac{1}{\varepsilon(\omega_{b},T)}=\frac{T}{T_{1}\varepsilon_{s1}%
+(T-T_{1})\varepsilon_{\infty}}\thickapprox\frac{T}{T_{1}\varepsilon_{s1}%
}.\label{die1}%
\end{equation}
The last step of Eq.(\ref{die1}) results from $\varepsilon_{s1}\gg
\varepsilon_{\infty}$ at all temperature. Unfortunately, no $\varepsilon
(\omega,T)$ data around $\omega_{b\text{ }}$are reported in this frequency
range \cite{lin}. Taking $T_{1}=$300K is convenient. Using the static
dielectric constant $\varepsilon_{0}\thickapprox70$ at $T_{1}=300$K
\cite{lin}, then $\varepsilon_{s1}\thicksim(\varepsilon_{\infty}%
+\varepsilon_{0})/2\thicksim38$, which is close to the interpolated value from
lower and higher frequencies, cf. Fig.2b of \cite{lin}. For $\varepsilon
_{\infty}=6.8,$ $\varepsilon_{0}=30$ \cite{bok}$,$ $\varepsilon_{s1}%
\thicksim(\varepsilon_{\infty}+\varepsilon_{0})/2\thicksim18$, which is quite
close to Re$\varepsilon(9$meV$)=20$ directly calculated from DFPT \cite{bok}.
Similarly, for $\varepsilon_{\infty}=4.5,$ $\varepsilon_{0}=24.1$
\cite{bri}$,$ $\varepsilon_{s1}\thicksim(\varepsilon_{\infty}+\varepsilon
_{0})/2\thicksim15$. As noticed in Sec.\ref{fc}, Eq.(\ref{die1}) are
applicable for $n_{e}<10^{18}$cm$^{-3}$. The effective interaction
$H_{\text{int}}$ relates to the bare interaction $H_{\text{int}}^{\text{bare}%
}$ by $H_{\text{int}}=H_{\text{int}}^{\text{bare}}/\varepsilon(\omega_{b},T)$,
where $H_{\text{int}}^{\text{bare}}$ represents any of $H_{\text{ee}%
}^{\text{bare}} $, $H_{\text{e-def}}^{\text{bare}}$ and $H_{\text{e-LA}%
}^{\text{bare}}$. Then,%

\begin{equation}
h_{\text{ee}}=\frac{e^{2}}{\epsilon_{0}\mathcal{V}|\mathbf{k}_{2}%
-\mathbf{k}_{2}^{\prime}|^{2}\varepsilon},\label{eee}%
\end{equation}
where $\varepsilon^{-1}$ is given by Eq.(\ref{die1}). The effective ee
interaction for whole sample is $H_{\text{ee}}=\sum_{j<k}h_{\text{ee}}(j,k)$.
Similarly,%
\begin{equation}
h_{\text{e-def}}=\frac{1}{4\pi\epsilon_{0}}\frac{\Delta z_{\text{I}^{-}}e^{2}%
}{\varepsilon R_{\text{e-V(I}^{-}\text{)}}},\label{eed}%
\end{equation}
the e-def interaction for whole sample is $H_{\text{e-def}}=\sum_{j\beta
}h_{\text{e-def}}(j,\beta)$.

\subsection{$h_{\text{P-LO}}$ and $h_{\text{P-LA}}$}

\label{pla}

The bare interaction of an electron with wave vector $\mathbf{k}_{i}$ with LA
phonons can be written as%
\[
h_{\text{e-LA}}^{\text{bare}}=-i\sum_{j\mathbf{Kk}}\frac{1}{\varepsilon}%
\sqrt{\frac{\hbar\mathcal{N}}{2M_{j}\omega_{\text{LA}}(\mathbf{k})}}%
\mathbf{e}_{j}^{(\text{LA})}(\mathbf{k})\cdot(\mathbf{k}+\mathbf{K})
\]%
\begin{equation}
V_{j\mathbf{k}+\mathbf{K}}c_{\mathbf{k}_{i}+\mathbf{k}+\mathbf{K}}^{\dag
}c_{\mathbf{k}_{i}}(a_{\mathbf{k}\text{LA}}+a_{-\mathbf{k}\text{LA}}^{\dag
}),\label{bea}%
\end{equation}
where $\omega_{\text{LA}}(\mathbf{k})=c_{l}k$, $c_{l}$ is the speed of
longitudinal sound wave. For a normal process ($\mathbf{K}=0$), one may
estimate Eq.(\ref{bea}) by%
\begin{equation}
h_{\text{e-LA}}^{\text{bare}}\thicksim\sum_{j\mathbf{k}}\sqrt{\frac
{\hbar\mathcal{N}}{2M_{j}c_{l}k}}k\frac{e^{2}}{\epsilon_{0}k^{2}\mathcal{V}%
}.\label{na}%
\end{equation}
Then the screened interaction is
\begin{equation}
h_{\text{e-LA}}\thicksim\sum_{\mathbf{k}}\frac{1}{\varepsilon(c_{l}k,T)}%
\sqrt{\frac{\hbar\mathcal{N}}{2M_{j}c_{l}k}}k\frac{e^{2}}{\epsilon_{0}%
k^{2}\mathcal{V}}.\label{ea1}%
\end{equation}
To simplify Eq.(\ref{na}), let us consider%
\[
\langle\frac{1}{k^{3/2}}\rangle=\frac{\int_{0}^{k_{b}}\frac{1}{k^{3/2}}4\pi
k^{2}dk}{\int_{0}^{k_{b}}4\pi k^{2}dk}=\frac{2}{k_{b}^{3/2}},
\]
then%
\[
\sum_{\mathbf{k}}\frac{1}{k^{3/2}}=\frac{\mathcal{V}\frac{4\pi}{3}k_{b}^{3}%
}{(2\pi)^{3}}\langle\frac{1}{k^{3/2}}\rangle=\frac{\mathcal{V}\frac{4\pi}%
{3}k_{b}^{3}}{(2\pi)^{3}}\frac{2}{k_{b}^{3/2}}.
\]
Eq.(\ref{na}) becomes:%
\[
h_{\text{e-LA}}^{\text{bare}}\thicksim\frac{\mathcal{V}\frac{4\pi}{3}k_{b}%
^{3}}{(2\pi)^{3}}\sqrt{\frac{\hbar\mathcal{N}}{2M_{j}c_{l}k_{b}}}2k_{b}%
^{-1}\frac{e^{2}}{\epsilon_{0}\mathcal{V}}.
\]

The interaction between an electron and LO phonons is \cite{kit}%

\begin{equation}
h_{\text{e-LO}}=\sum_{\mathbf{q}}(V_{\mathbf{q}}a_{\mathbf{q}}e^{i\mathbf{q}%
\cdot\mathbf{r}}+V_{\mathbf{q}}^{\dag}a_{\mathbf{q}}^{\dag}e^{-i\mathbf{q}%
\cdot\mathbf{r}}).\label{eo}%
\end{equation}
The order of magnitude of Eq.(\ref{eo}) is%
\begin{equation}
h_{\text{e-LO}}\thicksim\sum_{\mathbf{q}}i\hbar\omega_{\text{LO}}\frac
{1}{q\mathcal{V}^{1/2}}(\frac{\hbar}{2m\omega_{L}})^{1/4}[4\pi\alpha
]^{1/2},\label{eo1}%
\end{equation}
where%
\begin{equation}
\alpha=\frac{e^{2}}{4\pi\epsilon_{0}(2\hbar\omega_{\text{LO}})}(\frac
{2m\omega_{L}}{\hbar})^{1/2}(\frac{1}{\varepsilon_{\infty}}-\frac
{1}{\varepsilon_{0}}).\label{cls}%
\end{equation}
To further simplify Eq.(\ref{eo1}), let us consider%
\[
\langle\frac{1}{q}\rangle=\frac{\int_{0}^{k_{b}}\frac{1}{q}4\pi q^{2}dq}%
{\int_{0}^{k_{b}}4\pi q^{2}dq}=\frac{3}{2k_{b}},
\]
then%
\[
\sum_{\mathbf{q}}\frac{1}{q}=\frac{\mathcal{V}\frac{4\pi}{3}k_{b}^{3}}%
{(2\pi)^{3}}\langle\frac{1}{q}\rangle=\frac{\mathcal{V}\frac{4\pi}{3}k_{b}%
^{3}}{(2\pi)^{3}}\frac{3}{2k_{b}}.
\]
Eq.(\ref{eo1}) becomes%
\[
h_{\text{e-LO}}\thicksim\frac{\mathcal{V}\frac{4\pi}{3}k_{b}^{3}}{(2\pi)^{3}%
}\frac{3}{2k_{b}}\hbar\omega_{\text{LO}}\frac{1}{\mathcal{V}^{1/2}}%
(\frac{\hbar}{2m\omega_{L}})^{1/4}[4\pi\alpha]^{1/2}.
\]
Using $\mathcal{N}=n_{\text{cell}}\mathcal{V}$, where $n_{\text{cell}}$ is the
number of primitive cells in unit volume. One has%
\begin{equation}
h_{\text{e-LO}}/h_{\text{e-LA}}\thicksim\frac{\frac{3}{2}\hbar\omega
_{\text{LO}}(\frac{\hbar}{2m\omega_{L}})^{1/4}[4\pi\alpha]^{1/2}}{\frac
{1}{\varepsilon}\sqrt{\frac{\hbar}{2M_{j}c_{l}k_{b}}}2\frac{e^{2}}%
{\epsilon_{0}}n_{\text{cell}}^{1/2}},\label{oa}%
\end{equation}
where $\varepsilon^{-1}$ is given by Eq.(\ref{die1}). According to the polaron
theory \cite{feyp55}, the interaction between polaon and LO phonons (the
residual e-LO interaction) is small: $h_{\text{P-LO}}/h_{\text{e-LO}}%
\thicksim(\alpha/10)^{4}$, where $\alpha$ is the dimensionless coupling
constant given by Eq.(\ref{cls}). On the other hand, $h_{\text{P-Aph}%
}\thickapprox h_{\text{e-LA}}$. Finally,%
\begin{equation}
h_{\text{P-LO}}/h_{\text{P-LA}}\thicksim(\alpha/10)^{4}h_{\text{e-LO}%
}/h_{\text{e-LA}}\thicksim0.12.\label{ra}%
\end{equation}

If one use line width as an indication of the strength of e-Oph and e-Aph
interaction, then $h_{\text{e-LO}}\thicksim40$meV (MAPbI$_{3}$),
$h_{\text{e-LA}}\thicksim18$meV at T=300K \cite{wri}. It turns out
$h_{\text{P-LO}}/h_{\text{P-LA}}\thicksim(\alpha/10)^{4}h_{\text{e-:LO}%
}/h_{\text{e-LA}}\thicksim0.1$. Two different estimations produced similar
results. One can see that in MAPbI$_{3}$, P-LA interaction is more important
than P-LO interaction.

\subsection{Non-degenerate polaron gas}

Writing the average occupation number $f_{e}(\varepsilon)$ per spin in a state
with energy $\varepsilon$ as%
\begin{equation}
f_{e}(\varepsilon)=Ae^{-\varepsilon/k_{B}T},\label{bd2}%
\end{equation}
then the factor $A$ is determined by
\begin{equation}
n_{e}=2\int_{0}^{\infty}d\varepsilon f_{e}(\varepsilon)D(\varepsilon
),\label{den}%
\end{equation}
where%
\begin{equation}
D(\varepsilon)=\frac{1}{2\pi^{2}}\frac{m_{\text{P}}}{\hbar^{2}}\frac
{\sqrt{2m_{\text{P}}\varepsilon}}{\hbar}\label{dos}%
\end{equation}
is the number of states per unit energy per unit volume per spin. The energy
zero point is taken at $\varepsilon_{c}$, factor 2 in Eq.(\ref{den}) comes
from two spin states. Thus the average occupation number $f_{e}(\varepsilon) $
of a state with energy $\varepsilon$ is given by%
\begin{equation}
f_{e}(\varepsilon)=\frac{4\pi^{3/2}\hbar^{3}e^{-\varepsilon/k_{B}T}%
}{(2m_{\text{P}}k_{B}T)^{3/2}}n_{e}.\label{fd}%
\end{equation}

\begin{figure}
[H]%
\subfigure[]{\includegraphics[width=0.23\textwidth]{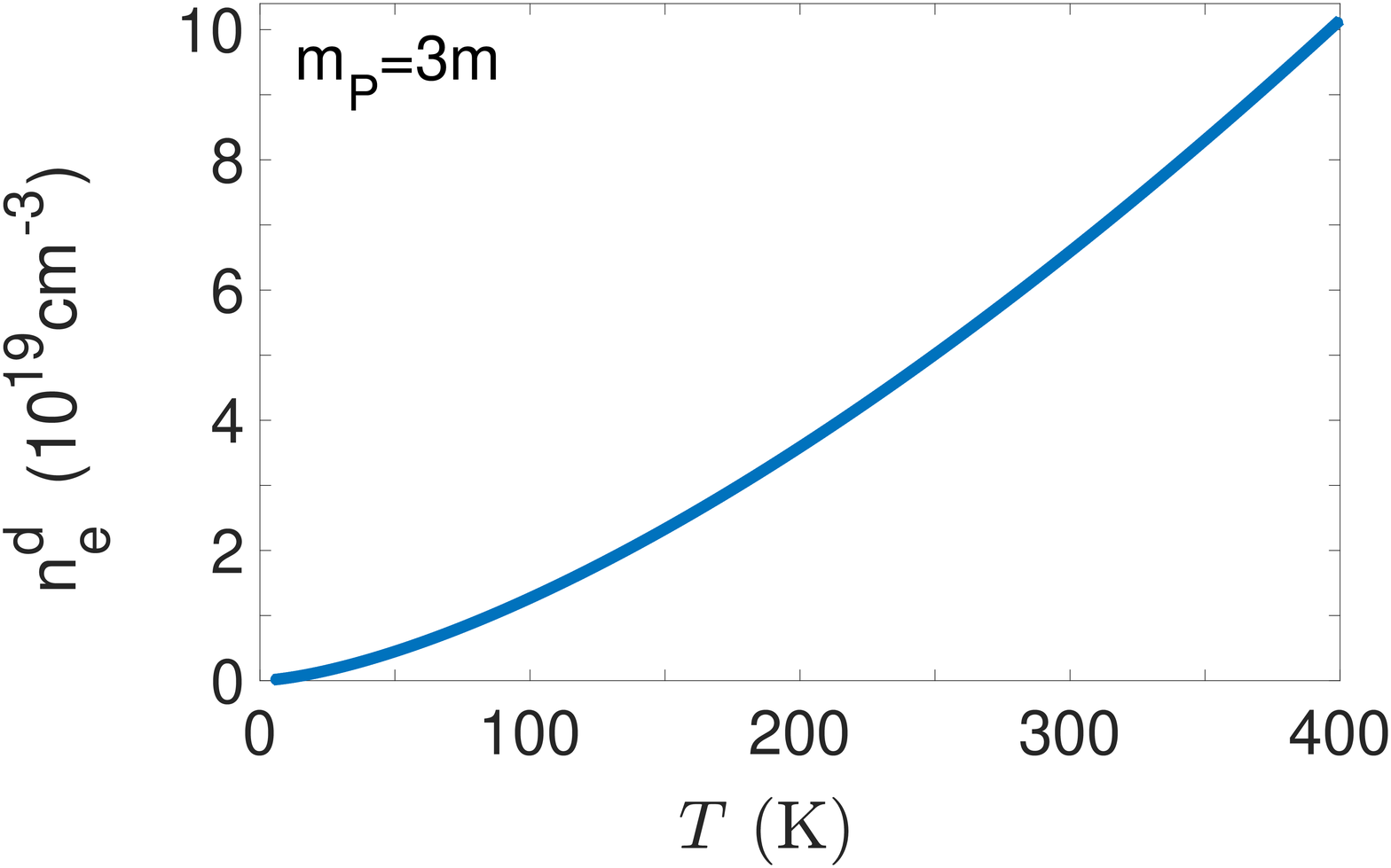}\label{den3}}
\subfigure[]{\includegraphics[width=0.23\textwidth]{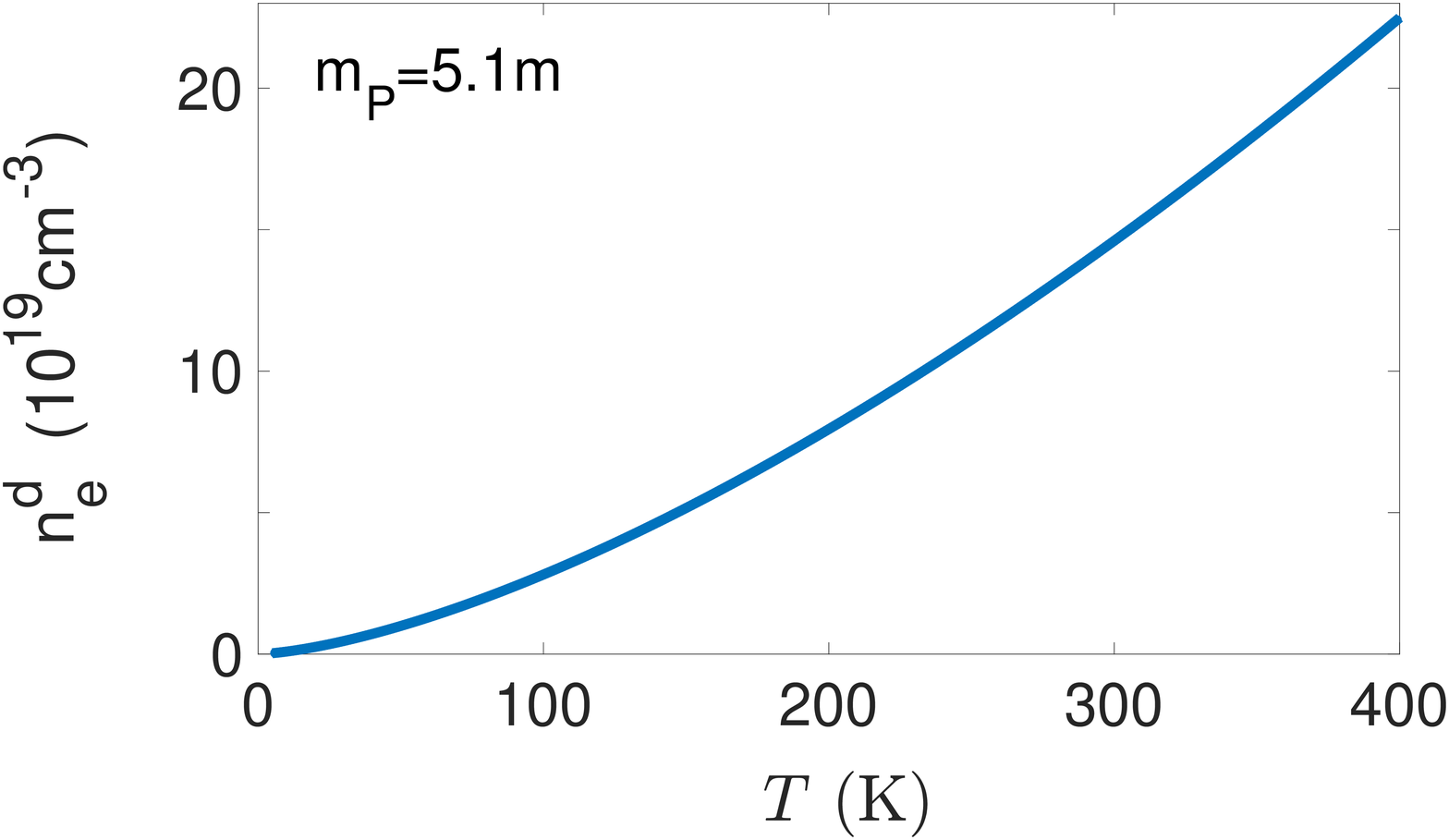}\label{den51}}
\subfigure[]{\includegraphics[width=0.23\textwidth]{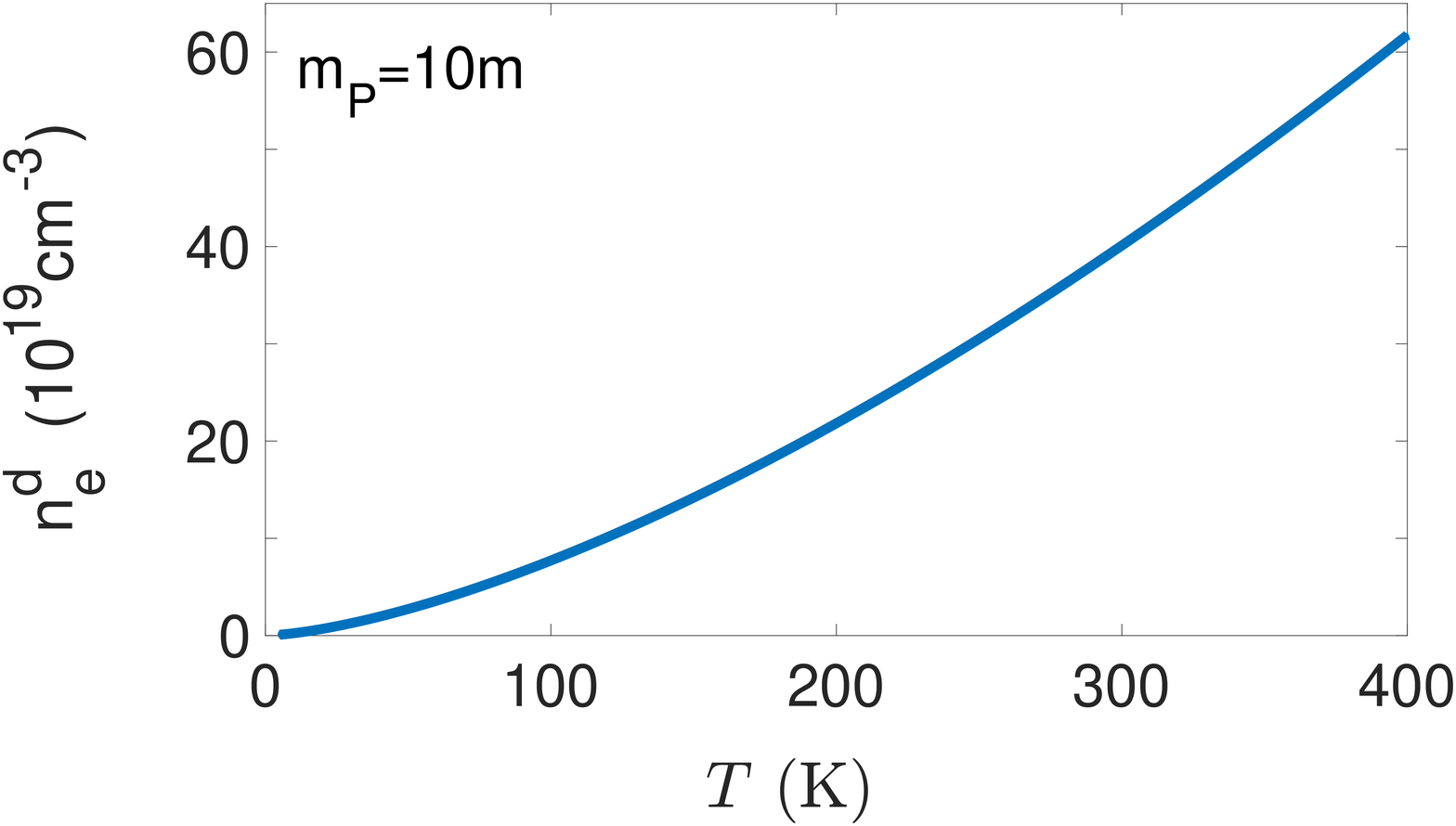}\label{den10}}
\subfigure[]{\includegraphics[width=0.23\textwidth]{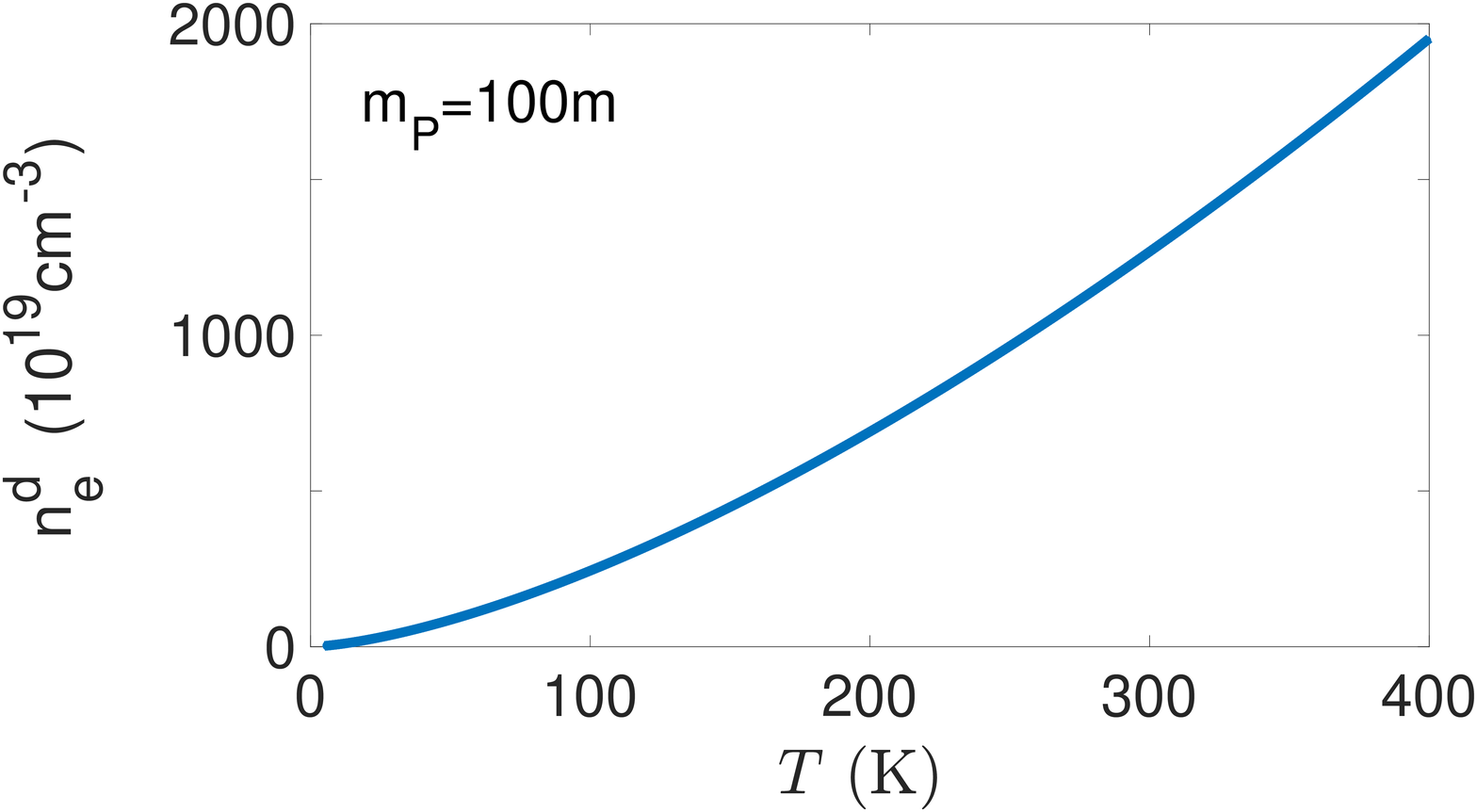}\label{den100}}
\caption{Degenerate density as function of temperature calculated from
Eq.(\ref{det}) for several values of $m_{\text{P}}$.} \label{dege}
\end{figure}

If we define the degeneracy of a gas of EPs as $e^{g/k_{B}T}=1$, the
degenerate density $n_{e}^{\text{d}}$ at a given temperature $T$ can be
determined by:%
\begin{equation}
n_{e}^{\text{d}}(T)=(m_{\text{P}}k_{B}T/2\pi\hbar^{2})^{3/2}.\label{det}%
\end{equation}
For several effective masses of polaron, Fig.\ref{dege} gives the degenerate
density $n_{e}^{\text{d}}$ as a function of temperature $T$.

The degenerate density $n_{e}^{\text{d}}\varpropto m_{\text{P}}^{3/2}$ is
sensitive to the choice of $m_{\text{P}}$. In Table.\ref{bad}, $n_{e}%
^{\text{d}}$ are listed for $T=10$, $20$ and $300$K. We could use Boltzmann
distribution (\ref{fd}) only when $n_{e}<(n_{e}^{\text{d}})_{\min}$. When
$n_{e}>n_{e}^{\text{d}}$, the gas of EPs or electrons should be described by
the Fermi distribution. In the photovoltaic application, $n_{e}\ll\min
\{n_{e}^{\text{d}},n_{e}^{\text{c}}\}$. The carriers are non-degenerate EP gas.

\begin{table}
[h]\caption {Degenerate density ($10^{19}$cm$^{-3}$) for several typical
temperature and polaron effective mass} \label{bad}
\begin{tabular}
[c]{|c|c|c|c|c|}\hline
T \textbackslash$m_{\text{P}}$ & 3m & 5.1m & 10m & 100m\\\hline
10K & 0.04 & 0.09 & 0.24 & 7.72\\\hline
20K & 0.11 & 0.25 & 0.69 & 21.83\\\hline
300K & 6.59 & 14.61 & 40.11 & 1268\\\hline
\end{tabular}

\end{table}

\subsection{Collision mechanisms}

\label{scm}

The formation energy $E_{\text{EP}}$ of an electron polaron (EP) is larger
than the maximal energy $\hbar\omega_{b}$ of an acoustic phonon (9.2meV), and
the energy change $v\Delta p\thicksim k_{B}T$ in EP-EP collision ($v$ is the
velocity of a polaron, $\Delta p$ is the change in momentum during a
collision). Therefore, before an EP annihilating with a hole polaron (HP), an
EP is a stable entity, will not break into an electron in three collision
mechanisms (EP-EP, EP-defect and EP-Aph).

The charge transport is mainly controlled by three collision mechanisms: (i)
EP-EP scattering; (ii) EP-defect scattering; and (iii) absorption or/and
emission a LA phonon by a EP. To calculate the mobility of EP with Boltzmann
equation, in a ABX$_{3}$ sample, let us consider a physical infinitesimal
volume \cite{v10} $\mathcal{V}$ which contains $\mathcal{N}$ primitive cells.

\subsection{EP-EP scattering}

\label{spp}

The rate $\nu_{\text{PP}}=(\frac{\partial f_{\mathbf{p}}}{\partial
t})_{\text{PP}}$ of change in distribution function $f_{\mathbf{p}}$ of EPs
caused by EP-EP collisions is \cite{pei}%
\[
(\frac{\partial f_{\mathbf{p}}}{\partial t})_{\text{PP}}=\frac{2\pi}{\hbar
}\sum_{\mathbf{p}_{2}\mathbf{p}^{\prime}}|\langle\mathbf{p}^{\prime
},\mathbf{p}_{2}^{\prime}|H_{\text{PP}}|\mathbf{p},\mathbf{p}_{2}\rangle|^{2}%
\]%
\begin{equation}
\delta(E_{\mathbf{p}}+E_{\mathbf{p}_{2}}-E_{\mathbf{p}^{\prime}}%
-E_{\mathbf{p}_{2}^{\prime}})\label{pe}%
\end{equation}%
\[
\{f_{\mathbf{p}^{\prime}}f_{\mathbf{p}_{2}^{\prime}}(1-f_{\mathbf{p}%
})(1-f_{\mathbf{p}_{2}})-f_{\mathbf{p}}f_{\mathbf{p}_{2}}(1-f_{\mathbf{p}%
^{\prime}})(1-f_{\mathbf{p}_{2}^{\prime}})\},
\]
where%
\begin{equation}
H_{\text{PP}}=\frac{e^{2}}{\epsilon_{0}\mathcal{V}|\mathbf{k}_{2}%
-\mathbf{k}_{2}^{\prime}|^{2}\varepsilon},\text{ }\label{eef}%
\end{equation}
is the effective Coulomb interaction between two EPs with momentum
$\mathbf{p}$ and $\mathbf{p}_{2}$, $\mathbf{p}_{2}-\mathbf{p}_{2}^{\prime
}=\mathbf{p}^{\prime}-\mathbf{p}$ is the momentum exchange during collision,
$\mathbf{k}_{2}-\mathbf{k}_{2}^{\prime}=(\mathbf{p}_{2}-\mathbf{p}_{2}%
^{\prime})/\hbar$. $\varepsilon^{-1}$ is given by Eq.(\ref{die1}).

To estimate the order of magnitude of $\nu_{\text{PP}}$, let us mimic the
procedure of \cite{pei}. Consider the second term in Eq.(\ref{pe}). We notice
that $f_{\mathbf{p}}=1$, $f_{\mathbf{p}_{2}}$ is given by Eq.(\ref{fd}) by
$\varepsilon\rightarrow\varepsilon_{\mathbf{p}_{2}}$. Because $f_{\mathbf{p}%
^{\prime}}$ and $f_{\mathbf{p}_{2}^{\prime}}\ll1$, the number in each\ ( ) can
be taken as 1. If one views the energy conservation delta function as a
rectangle with width $D$, where $D\thicksim3$eV is the width of the conduction
band \cite{bri,pro}. Then the height of the delta function is $D^{-1}$. The
summations over $\mathbf{p}_{2}$ and $\mathbf{p}^{\prime}$ produces
$(k_{B}T\mathcal{V}\frac{dZ}{dE})^{2}$, here $\frac{dZ}{dE}\thicksim
a^{-3}D^{-1}$ is the density of states per unit volume per unit energy
interval, $a$ is the lattice constant of a primitive cell. The characteristic
momentum exchange may be taken as the minimal detectable change in momentum,
i.e. the uncertainty of the momentum $\hbar/d$, where $d=2R_{\text{P}}$ is the
diameter of a EP. Then the average collision frequency $\nu_{\text{PP}}$ of a
EP with other EPs is
\begin{equation}
\nu_{\text{PP}}\thicksim\left[  \frac{T}{T_{1}\varepsilon_{s1}+(T-T_{1}%
)\varepsilon_{\infty}}\right]  ^{2}n_{e}\label{ee}%
\end{equation}%
\[
\frac{4\pi^{3/2}\hbar^{3}e^{-\varepsilon_{2}/k_{B}T}}{(2m_{EP}k_{B}T)^{3/2}%
}\frac{1}{\hbar}\frac{d^{4}}{a^{6}}(\frac{e^{2}}{\epsilon_{0}})^{2}%
\frac{(kT)^{2}}{D^{3}}\text{.}%
\]
Because the EP gas is non-degenerate, $n_{e}$ and $m_{\text{P}}^{-3/2}$ appear
in $f_{\mathbf{p}_{2}}$, and as a result appear in $\nu_{\text{PP}}$.
According to the energy partition theorem, the average energy $\varepsilon
_{2}\thicksim3k_{B}T/2$, then $e^{-\varepsilon_{2}/k_{B}T}\thickapprox
e^{-3/2}$ does almost not depend on temperature. If EP-EP collision was the
unique collision mechanism, the mobility $\mu=e/\nu_{\text{PP}}m_{EP}$ of EP
would be proportional to $n_{e}^{-1}m_{\text{P}}^{1/2}T^{-5/2}$.

\subsection{Scattering EP by defects}

\label{spv}Denote $C$ as the number of the I$^{-}$ vacancies per unit volume,
the average distance between two vacancies is $\thicksim C^{-1/3}$. The
Coulomb scattering amplitude caused by $h_{\text{def}}$ is \cite{v3,scf}
$f_{c}\thicksim\Delta z_{\text{I}^{-}}e^{2}/8\pi\epsilon_{0}\varepsilon
k_{B}T\thickapprox1.76$\AA , is smaller than $C^{-1/3}$ for a moderate defect
concentration. For example, if the concentration of I$^{-}$ vacancy is 1 in 2
cells ($16.7\%$), the average distance between two I$^{-}$ vacancies is
$\thicksim7.9$\AA . Then one can ignore the interference between two I$^{-}$
vacancies. In a sample with volume $\mathcal{V}$, there are $C\mathcal{V}$
iodine ions vacancies. The total scattering probability with $C\mathcal{V}$
I$^{-}$ vacancies is just $C\mathcal{V}$ multiply by the scattering
probability with one I$^{-}$ vacancy.

Comparing to the kinetic energy of an EP, even taken into account the large
dielectric constant $\varepsilon\thickapprox70$ of CH$_{3}$NH$_{3}$PbI$_{3} $,
the polaron-defect interaction $h_{\text{P-def}}$ cannot be treated as a
perturbation \cite{cal,v3}, Born approximation is inapplicable \cite{v3,v10}
to.the scattering of an EP by an I$^{-}$ vacancy. Fortunately, for Coulomb
potential (\ref{eed}), the scattering probability calculated by Fermi's golden
rule is the same as that calculated by the exact method \cite{bohm}.

The rate of change in the distribution function of EPs caused by the
collisions with I$^{-}$ vacancies is \cite{v10,cal}%
\begin{equation}
(\frac{\partial f_{\mathbf{p}}}{\partial t})_{\text{P-def}}=\sum
_{\mathbf{p}^{\prime}}C\mathcal{V}\frac{2\pi}{\hbar}|\langle\mathbf{p}%
^{\prime}|h_{\text{P-def}}|\mathbf{p}\rangle|^{2}\label{cdf}%
\end{equation}%
\[
\delta(E_{\mathbf{p}^{\prime}}-E_{\mathbf{p}})[f_{\mathbf{p}^{\prime}%
}(1-f_{\mathbf{p}})-f_{\mathbf{p}}(1-f_{\mathbf{p}^{\prime}})],
\]
where%
\begin{equation}
\langle\mathbf{p}^{\prime}|h_{\text{P-def}}|\mathbf{p}\rangle=\frac
{1}{\varepsilon\mathcal{V}\epsilon_{0}}\frac{\Delta z_{\text{I}^{-}}e^{2}%
}{q^{2}},\text{ }\mathbf{q}=(\mathbf{p}^{\prime}-\mathbf{p})/\hbar.\label{cm}%
\end{equation}

Let us estimate the average collision frequency $\nu_{\text{P-def}}=(\partial
f_{\mathbf{p}}/\partial t)_{\text{P-def}}$ of an EP with defects. Since an
I$^{-}$ vacancy is attached to the perovskite lattice, and the mass of lattice
is much larger than $m_{\text{P}}$. The collision of an EP with an I$^{-}$
vacancy can be viewed as elastic collision, a typical change in wave vector is
$q\thicksim\sqrt{2m_{\text{P}}k_{B}T}/\hbar$. $\sum_{\mathbf{p}^{\prime}}$ is
over $\mathcal{V}k_{B}TD^{-1}a^{-3}$ electronic states. Viewing the delta
function as a rectangle with width $D$, the height of delta function is
$\thicksim D^{-1}$. Combine these factors, the rate $\nu_{\text{P-def}}$ of
change in the distribution function of EPs caused by the collisions with
I$^{-}$ vacancies is%
\begin{equation}
\nu_{\text{P-def}}\thickapprox\left[  \frac{T}{T_{1}\varepsilon_{s1}%
+(T-T_{1})\varepsilon_{\infty}}\right]  ^{2}C\frac{2\pi}{\hbar}\label{cdf1}%
\end{equation}%
\[
\left(  \frac{\Delta z_{\text{I}^{-}}e^{2}}{\epsilon_{0}}\right)  ^{2}\frac
{1}{D^{2}a^{3}}\frac{\hbar^{4}}{(2m_{\text{P}}k_{B}T)^{2}}k_{B}T.
\]
The circle line in Fig.\ref{vaca} is $\nu_{\text{P-def}}$ vs. $T$ for a
moderate vacancy concentration $C=1/10$cell ($3.33\%$). $\nu_{\text{P-def}}$
is much smaller than $\nu_{\text{P-Aph}}$. However, for $C\thicksim1/$cell
$(33.3\%)$, $\nu_{\text{P-def}}$ is comparable to even surpass $\nu
_{\text{P-Aph}}$.

It has been noticed that $\mu$ is not sensitive to the defects in ABX$_{3}$
\cite{zhu,bre}. One of the reasons is that ABX$_{3}$ is in a super
paraelectric phase, has a larger dielectric function. The polaron-defect
interaction is screened by the internal electric field caused by the
displacements of the Pb$^{2+}$ and I$^{-}$ ions: $h_{\text{P-def}%
}=h_{\text{P-def}}^{\text{bare}}/\varepsilon\ll h_{\text{P-def}}^{\text{bare}%
}$. According to Eq.(\ref{cdf1}), $\nu_{\text{P-def}}\thicksim C/\varepsilon
^{2}$. If ABX$_{3}$ was not in a super paraelectric phase, $\varepsilon$ would
be order 1. The polaron-defect scattering would be a far more serious matter.

\subsection{Absorption or emission a LA phonon by a EP}

\label{spap}

The rate $\nu_{\text{P-LA}}=(\partial f_{\mathbf{p}}/\partial t)_{\text{P-LA}%
}$ of change in the distribution function $f_{\mathbf{p}}$ of EP caused by the
P-LA phonon scattering is \cite{v10}%

\begin{equation}
(\frac{\partial f_{\mathbf{p}}}{\partial t})_{\text{P-LA}}=-\sum_{\mathbf{k}%
}\frac{\partial N_{0}(\omega)}{\partial\hbar\omega}[f_{0}(\mathbf{p}^{\prime
})-f_{0}(\mathbf{p})]\label{ep1}%
\end{equation}%
\[
\{w(\mathbf{p}^{\prime},\mathbf{k};\mathbf{p})(\varphi_{\mathbf{p}^{\prime}%
}-\varphi_{\mathbf{p}}+\chi_{\mathbf{k}})\delta(E_{\mathbf{p}}-E_{\mathbf{p}%
^{\prime}}-\hbar\omega_{\mathbf{k}})
\]%
\[
-w(\mathbf{p}^{\prime};\mathbf{p},\mathbf{k})(\varphi_{\mathbf{p}^{\prime}%
}-\varphi_{\mathbf{p}}-\chi_{\mathbf{k}})\delta(E_{\mathbf{p}}-E_{\mathbf{p}%
^{\prime}}+\hbar\omega_{\mathbf{k}})\}
\]
where $f_{0}$ and $N_{0}$ are the equilibrium distribution functions at
temperature $T$ for EPs and phonons, $\varphi$ and $\chi$ describe the
deviations of $f_{\mathbf{p}}$ and $N_{\mathbf{k}}$ from equilibrium%
\begin{equation}
f_{\mathbf{p}}-f_{0}(\varepsilon)=-\frac{\partial f_{0}(\varepsilon)}%
{\partial\varepsilon}\varphi,\label{df}%
\end{equation}
and%
\begin{equation}
N_{\mathbf{k}}-N_{0}(\mathbf{k})=-\frac{\partial N_{0}(\omega)}{\partial
\hbar\omega}\chi.\label{dn}%
\end{equation}
The probability coefficient $w(\mathbf{p}^{\prime},\mathbf{k};\mathbf{p})$ is
defined by%
\begin{equation}
\frac{2\pi}{\hbar}|\langle\mathbf{p}^{\prime},\mathbf{k}|h_{\text{P-LA}%
}(\text{emission})|\mathbf{p}\rangle|^{2}\label{w1}%
\end{equation}%
\[
=w(\mathbf{p}^{\prime},\mathbf{k};\mathbf{p})(N_{\mathbf{k}}+1),
\]
or%
\[
\frac{2\pi}{\hbar}|\langle\mathbf{p}^{\prime}|h_{\text{P-LA}}%
(\text{absorption})|\mathbf{p},\mathbf{k}\rangle|^{2}%
\]%
\begin{equation}
=w(\mathbf{p}^{\prime};\mathbf{p},\mathbf{k})N_{\mathbf{k}}.\label{w2}%
\end{equation}
Using Eqs.(\ref{ef},\ref{w1}), one has%
\begin{equation}
w\thicksim\frac{\pi\mathcal{N}}{\varepsilon^{2}M\omega k^{2}\mathcal{V}^{2}%
}(\frac{ze^{2}}{\epsilon_{0}})^{2}.\label{w}%
\end{equation}
In calculating transition probability caused by $h_{\text{P-LA}}$, one should
be careful that the total probability is NOT the sum of the probabilities
caused by the ions in a primitive cell. The interference between ions is vital
\cite{c60}, otherwise will lead to a wrong conclusion: $w=\sum_{i}w_{i}$ which
means that the more atoms in a primitive cell, the stronger P-LA interaction.

Let us estimate the change rate $\nu_{\text{P-LA}}=(\frac{\partial
f_{\mathbf{p}}}{\partial t})_{\text{P-LA}}$ of distribution function of EP
caused by emission or absorption a LA phonon. $\sum_{\mathbf{k}}$ in
Eq.(\ref{ep1}) represents the summation over all possible phonon states,
produces a factor $\mathcal{V}4\pi k_{b}^{3}/3$. The delta function represents
the energy conservation during the emission or absorption of a phonon. Since
the maximal allowable energy of a phonon is $\hbar\omega_{b}$, the width of
the delta function is $\hbar\omega_{b}$. Then the height of delta function is
$(\hbar\omega_{b})^{-1}$. For most of the temperature range of photovoltaic
application, $\hbar\omega_{b}\ll k_{B}T$, so that \cite{v10}
\begin{equation}
\frac{\partial N_{0}(\omega)}{\partial\hbar\omega}\thicksim\frac{k_{B}%
T}{(\hbar\omega_{b})^{2}}.\label{bs}%
\end{equation}
Because the characteristic energy of an EP is $\thicksim k_{B}T$ and
$\hbar\omega_{b}\ll k_{B}T$, $(\varphi_{\mathbf{p}^{\prime}}-\varphi
_{\mathbf{p}}+\chi_{\mathbf{k}})\thicksim k_{B}T$ \cite{v10}. The delta
function in Eq.(\ref{ep1}) requires $E_{\mathbf{p}^{\prime}}=E_{\mathbf{p}}%
\pm\hbar\omega_{\mathbf{k}}$, then
\[
f_{0}(\mathbf{p}^{\prime})-f_{0}(\mathbf{p})=\frac{\partial f_{0}%
(\varepsilon)}{\partial\varepsilon}(E_{\mathbf{p}^{\prime}}-E_{\mathbf{p}})
\]%
\begin{equation}
\thicksim n_{e}\frac{4\pi^{3/2}\hbar^{3}e^{-\varepsilon/k_{B}T}}%
{(2m_{\text{P}})^{3/2}(k_{B}T)^{5/2}}\hbar\omega_{b}.\label{fer}%
\end{equation}
Since we are considering the change of occupation number in state
$|\mathbf{p}\rangle$ at the time moment $t$, in Eq.(\ref{ep1}), $f_{\mathbf{p}%
}=1$ at the concerned moment $t$ \cite{pei}. Combine above considerations, one
has%
\[
\nu_{\text{P-LA}}\thicksim\left[  \frac{T}{T_{1}\varepsilon_{s1}%
+(T-T_{1})\varepsilon_{\infty}}\right]  ^{2}\frac{\pi}{M\omega_{b}k_{b}%
^{2}a^{3}}%
\]%
\begin{equation}
\frac{4\pi}{3}k_{b}^{3}(\frac{ze^{2}}{\epsilon_{0}})^{2}(\frac{k_{B}T}%
{\hbar\omega_{b}})^{2}n_{e}\frac{4\pi^{3/2}\hbar^{3}e^{-\varepsilon/k_{B}T}%
}{(2m_{EP})^{3/2}(k_{B}T)^{5/2}}.\label{cip}%
\end{equation}
In Fig.\ref{EPC}, Eqs.(\ref{ee},\ref{cdf1},\ref{cip}) are plotted\ against $T
$ for $n_{e}=10^{18}$cm$^{-1}$, $m_{\text{P}}=10m$, $\varepsilon(\omega
_{b})=36.5$ [which are different to those for Fig.3 in the text]. One can see
that relation $\nu_{\text{P-LA}}\gg\nu_{\text{P-def}}\gg\nu_{\text{PP}}$ is
not sensitive to the choice of $m_{\text{P}}$ and $\varepsilon$. One can also
see $\nu_{\text{P-Aph}}\gg\nu_{\text{P-def}}$ intuitively: (1) the available
states for the EP-defect scattering produce a factor $1/D^{2}$, while for the
P-Aph process produce a factor $(\hbar\omega_{D})^{-2}$, $1/D^{2}\ll
(\hbar\omega_{D})^{-2}$; (2) the probability of P-Aph interaction is
proportional to $N_{\omega}$ which is a large number because the acoustic
modes are fully excited.

We calculate $\mu(T)$\ with the experimental $\varepsilon(\omega,T)$
for\ $\omega/2\pi=1$KHz obtained in \cite{ono} instead of $\varepsilon
_{\infty}+C(\omega)/T$, i.e. in Eqs.(\ref{ee},\ref{cdf1},\ref{cip}) replace
$\left[  \frac{T}{300\varepsilon_{s1}+(T-300)\varepsilon_{\infty}}\right]
^{2}$\ by $[\varepsilon(\omega_{b},T)]^{-2}$, where $\varepsilon(\omega
_{b},T)=[\varepsilon(\omega/2\pi=1$KHz$,T)+\varepsilon_{\infty}]/2$. The
general trend of $\mu(T)$\ observed in [\onlinecite{hut}] is reproduced: in
tetragonal phase $\mu(T)\varpropto T^{-3/2}$, while in the orthorhombic phase
$\mu(T)$\ decreases with decreasing temperature. Due to the samples in
[\onlinecite{hut}] \ and in [\onlinecite{ono}] are different, the transition
temperature in [\onlinecite{hut}] is 150K, while in [\onlinecite{ono}] is
160K. But the general trend of $\mu(T)$\ in orthorhombic phase \cite{hut} is
reproduced by the current model with experimental dielectric function\ of
[\onlinecite{ono}]. If one assumes that (i) the sample in [\onlinecite{hut}]
is more uniform and is annealed slowly, below 150K, only orthorhombic phase
exists; (ii) both tetragonal phase and orthorhombic phase coexist
\cite{kong,hut} in the samples of earlier work \cite{oga,sav,mil} at $T<150$K;
then the two observations can be conciliated.

\begin{figure}
[H]%
\centering\subfigure[]{\includegraphics[width=0.23\textwidth]{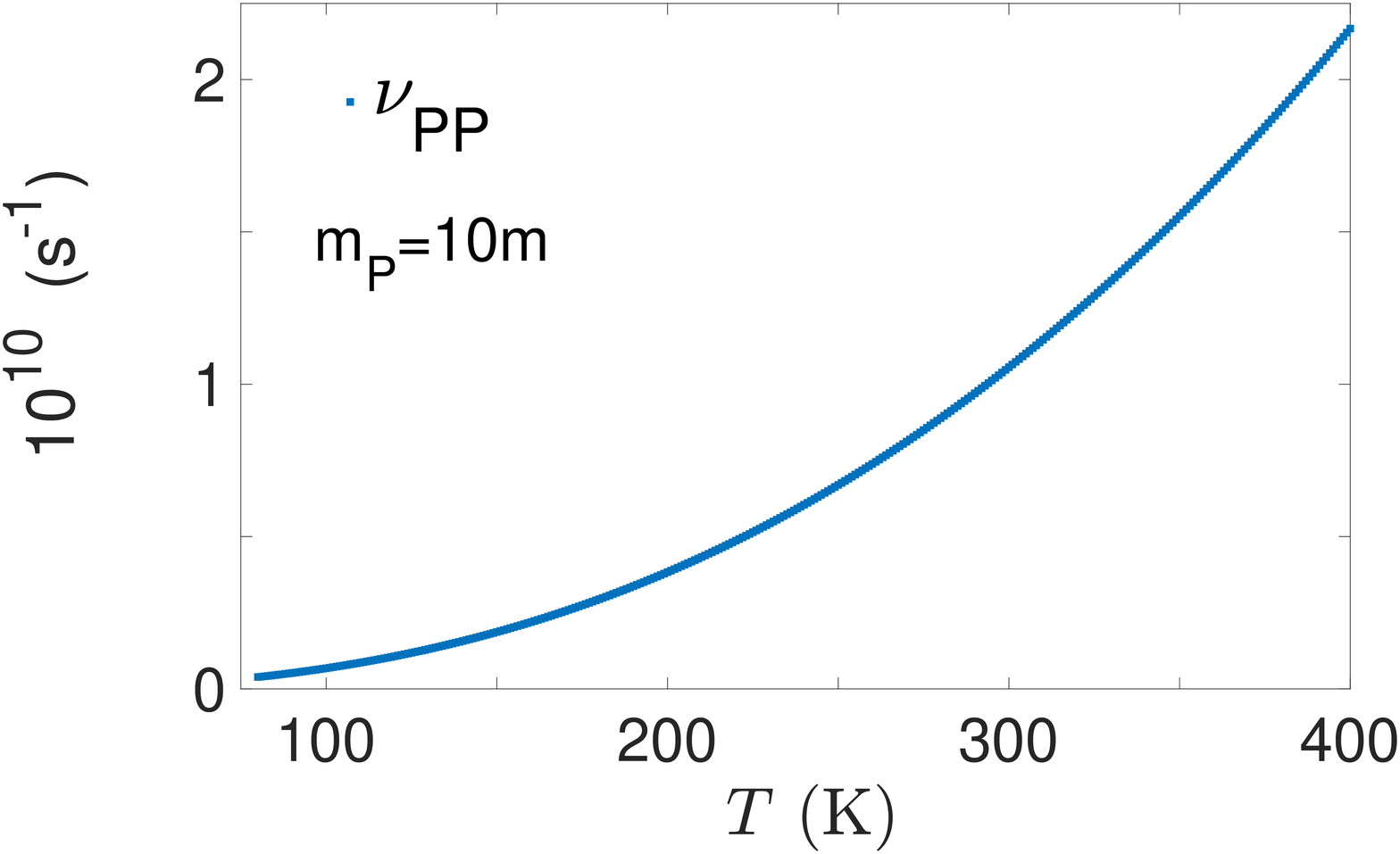}\label{eeP}}
\subfigure[]{\includegraphics[width=0.23\textwidth]{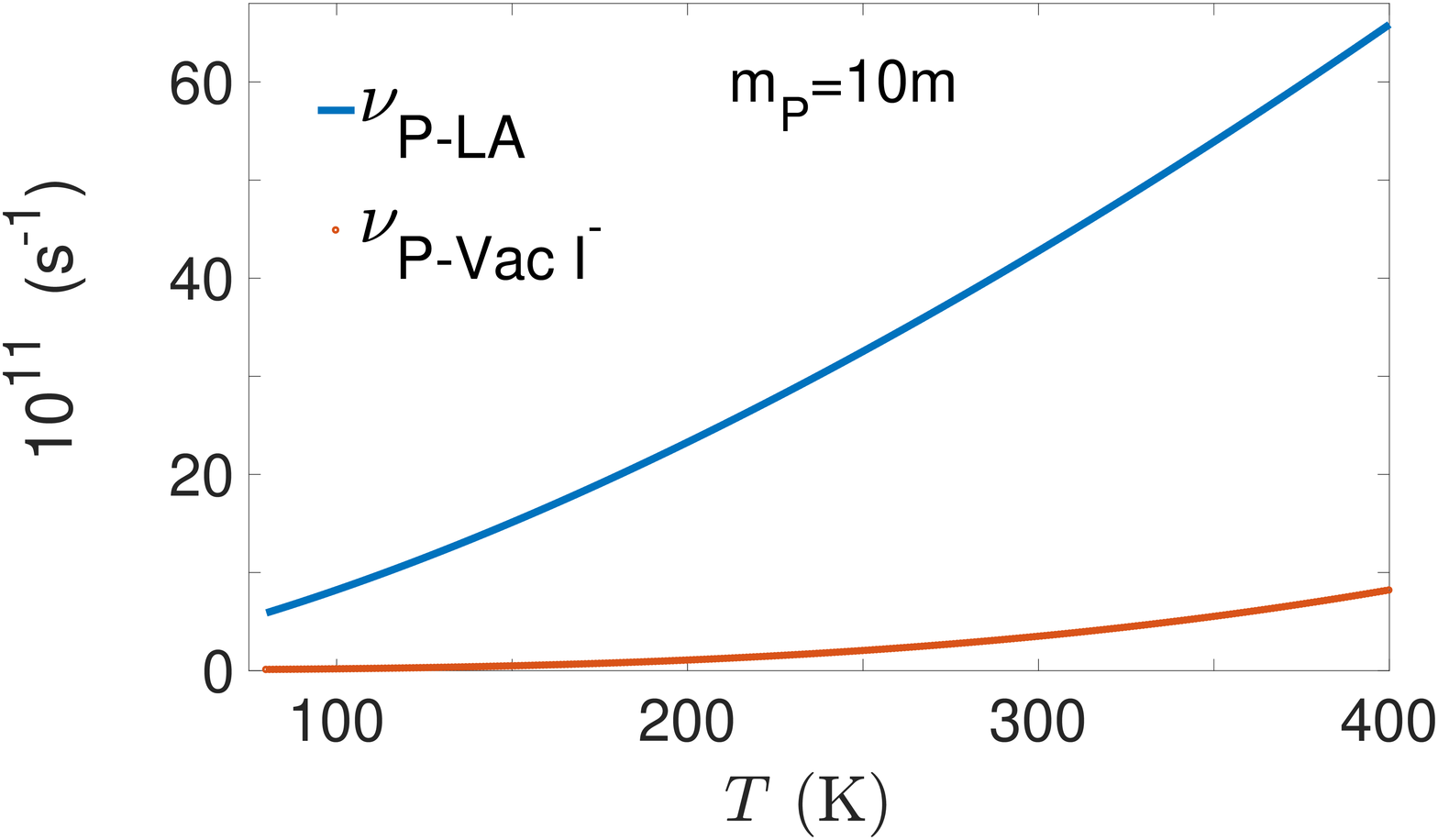}\label{vaca}}
\caption{$m_{\text{P}}=10m$, $n_{e}=10^{18}$cm$^{-3}$, $c_{s}=2147$ m/s, using last step of Eq.(\ref{die1}) and $\varepsilon(\omega_{0})=36.5$. (a) PP
collision frequency $\nu_{\text {PP}}$ as a function of temperature
determined by Eq.(\ref{ee}). (b) P-LA collision frequency
$\nu_{\text{P-LA}}$ and P-Vac collision frequency $\nu_{\text{P-Vac
I$^{-}$}}$ as functions of temperature determined by Eq.(\ref{cip}) and
Eq.(\ref{cdf1}). $\nu_{\text{P-LA}}$ is the main scattering mechanism.}
\label{EPC}
\end{figure}

The total collision frequency $\nu_{t}$ of an EP is the summation of all three
processes:
\begin{equation}
\nu_{t}=\nu_{\text{P-LA}}+\nu_{\text{P-def}}+\nu_{\text{PP}}.\label{tot}%
\end{equation}
The mobility is%
\begin{equation}
\mu=\frac{e}{\nu_{t}m_{\text{P}}}.\label{mo}%
\end{equation}

\begin{figure}
[H]\centering
\subfigure[]{\includegraphics[width=0.23\textwidth]{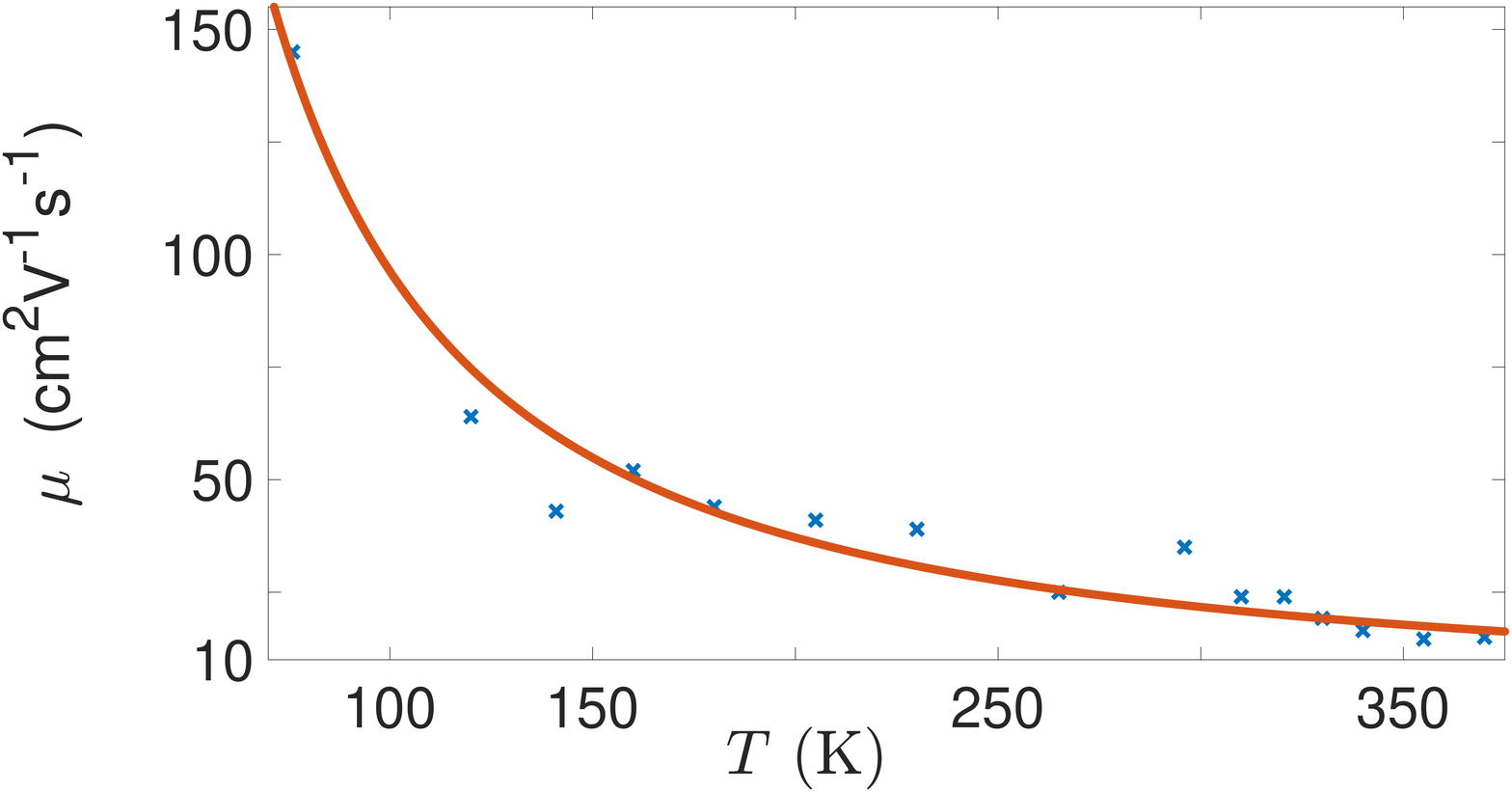}\label{milExa70}}
\subfigure[]{\includegraphics[width=0.23\textwidth]{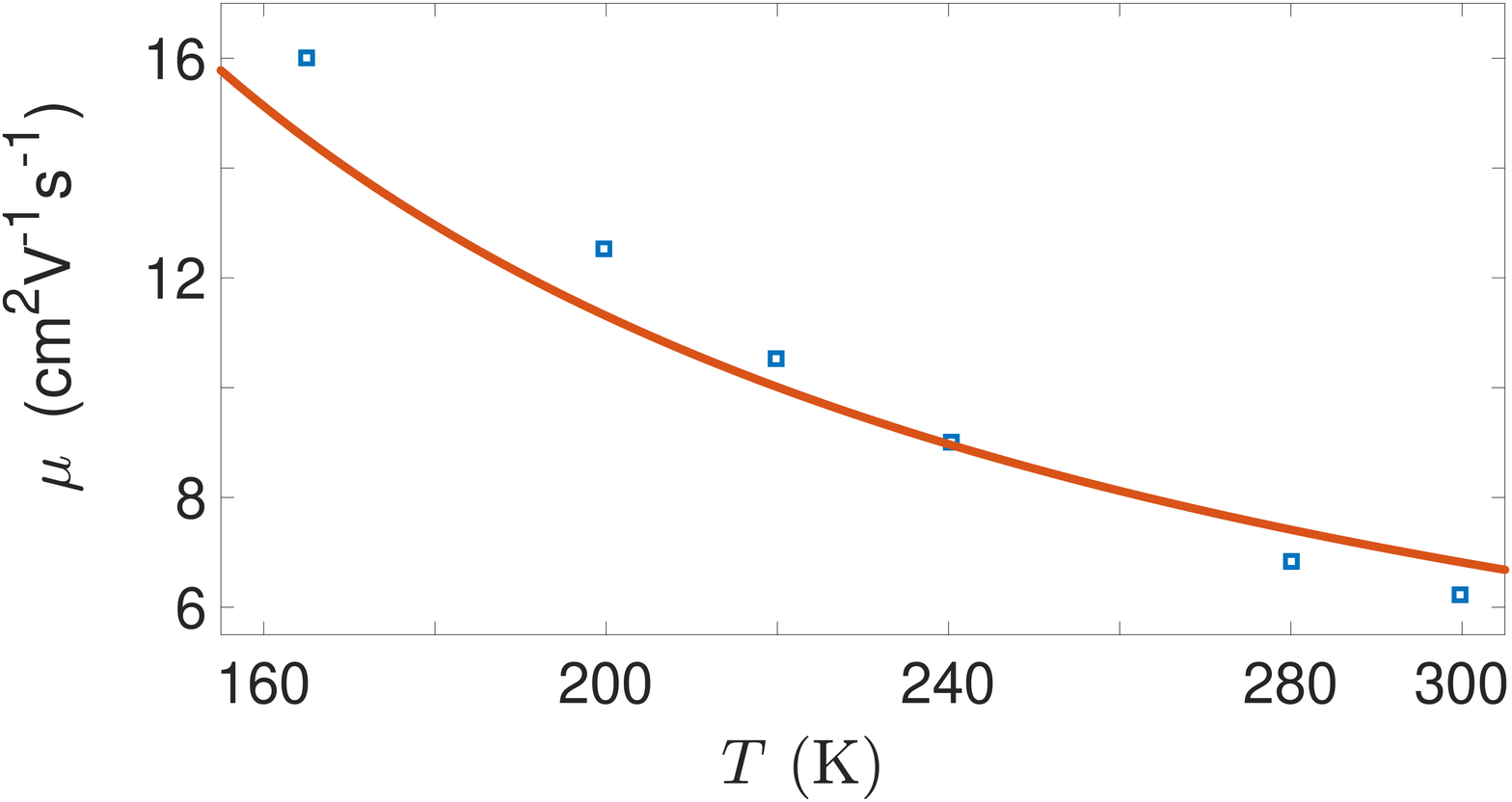}
\label{SavExa70}}
\caption{Mobility calculated with Eqs.(\ref{ee},\ref{cdf1},\ref{cip},\ref{tot},\ref{mo})(solid line) and exact Eq.(\ref{die1}), dielectric function $\varepsilon_{0}=70$, $\varepsilon_{\infty}=6.5$ taken from measurement \cite{lin}.
(a) Experimental data (crosses) from \cite{mil}, $n_{e}=1.2\times 10^{18}$cm$^{-3}$. (b) Experimental
data (squares) from \cite{sav}, $n_{e}=4.1\times 10^{18}$cm$^{-3}$.}\label{mob70}%

\end{figure}

From Eqs.(\ref{ee},\ref{cdf1},\ref{cip}), we have seen that collision
frequencies and $\mu$ depend on $\varepsilon(\omega,T)$. There are some
discrepancies among the experimental dielectric functions obtained by
different authors. For example, $\varepsilon_{0}\thickapprox33$ according to
\cite{pog,fro14} instead of $\varepsilon_{0}\thickapprox70$ \cite{lin,ono},
that may due to the differences in the compositions, structures, status of
samples and the methods of measurement. The dielectric functions calculated by
different methods are not far from each other $\varepsilon_{0}\thickapprox
24.5$, about two times smaller than the well cited experimental value
$\varepsilon_{0}\thickapprox70$ \cite{lin,ono,sew,sewpr}. In Fig.\ref{mob70},
we made a fitting with $\varepsilon_{0}\thickapprox70 $, which anticipates a
higher concentration of carriers. However, Eq.(\ref{mo}) is only for the
mobility $\mu_{e}$ of EPs. The hole mobility $\mu_{h} $ has the same order of
magnitude as $\mu_{e}$. The measured mobility is $\phi(\mu_{e}+\mu_{h})$,
where $\phi$ is the quantum efficiency of photon. If we plot $(\mu_{e}+\mu
_{h})$ rather than just $\mu_{e}$, $\varepsilon_{0}\thickapprox70$ would
suggest a similar carrier concentration as that obtained in Fig.5 in text.

\subsection{Decreases of vibrational entropy in a polaron}

The frequency $\omega_{1}$ of Pb-I stretch mode is increased in a polaron
which leads to the decrease of vibrational entropy. To estimate the decrease
of vibrational entropy, we apply the Morse potential%
\begin{equation}
V(r)=D[1-e^{-\sigma(r-r_{e})}]^{2},\label{mor}%
\end{equation}
to describe the potential energy of the Pb$^{2+}$ and I$^{-}$ interaction,
where $\sigma$ is a positive number, $D$ is the dissociation energy of Pb-I
bond, $r$ is the distance between Pb$^{2+}$ and I$^{-}$ ion, $r_{e}%
=3.15$\AA \ is the equilibrium distance between Pb$^{2+}$ and I$^{-}$ in a
normal lattice without the static deformation inside an EP.

\begin{figure}
[ht]\centering{\includegraphics[width=0.4\textwidth]{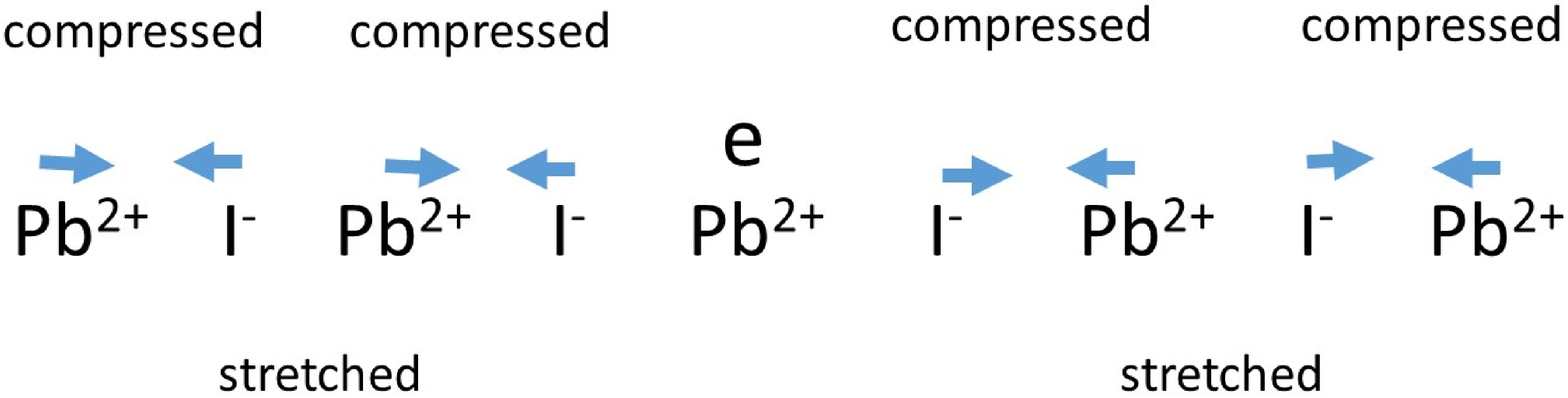}}
\caption{Schematic sketch of the deformation in an EP: Pb-I bonds are
compressed and stretched alternatively by the extra electron in the center
of polaron.} \label{poldef}
\end{figure}

In an EP, Pb-I bonds are alternatively compressed and stretched, cf.
Fig.\ref{poldef}. These static deformations make the Pb-I distances are
smaller and larger than $r_{e}$. If the distance between Pb and I is fixed at
$r$, the spring constant $k(r)$ is%
\begin{equation}
k(r)=\frac{d^{2}V(r)}{dr^{2}}=2D\sigma^{2}[2e^{-2\sigma(r-r_{e})}%
-e^{-\sigma(r-r_{e})}].\label{sk}%
\end{equation}
The positive number $\sigma$ in Eq.(\ref{mor}) is determined by%
\begin{equation}
\sigma=\sqrt{\frac{k(r_{e})}{2D}}.\label{sig}%
\end{equation}
Then, Eq.(\ref{sk})\ becomes%
\begin{equation}
k(r)=k(r_{e})[2e^{-2\sigma(r-r_{e})}-e^{-\sigma(r-r_{e})}].\label{sk1}%
\end{equation}

One can easily check%
\begin{equation}
\frac{dk(r)}{dr}=2D\sigma^{3}[e^{-\sigma(r-r_{e})}-4e^{-2\sigma(r-r_{e}%
)}],\label{kd}%
\end{equation}
and
\begin{equation}
\frac{d^{2}k(r)}{dr^{2}}=2D\sigma^{4}[8e^{-2\sigma(r-r_{e})}-e^{-\sigma
(r-r_{e})}].\label{kd2}%
\end{equation}
From Eq.(\ref{kd}), one can see $\frac{dk(r)}{dr}|_{r=r_{e}}=0$. From
Eq.(\ref{kd2}), one can see $\frac{d^{2}k(r)}{dr^{2}}>0$ for
\begin{equation}
r<r_{e}+\sqrt{\frac{2D}{k(r_{e})}}\ln8.\label{str}%
\end{equation}
Above two facts mean that $r=r_{e}$ is a minimum of $k(r)$. The spring
constant of the stretch mode of Pb-I bond can be estimated from $k(r_{e}%
)=m_{r}\omega_{\text{LO}}^{2}$, where $m_{r}$ is the reduced mass of Pb ion
and I ion. If one takes $\omega_{\text{LO}}=100$cm$^{-1}$, $k(r_{e})=46.4$N/m.
The bond energy of Pb-I is $D=142$KJ/mol [http://www.chem.tamu.edu/rgroup

/connell/linkfiles/bonds.pdf], then $\sqrt{\frac{2D}{k(r_{e})}}\ln8=2.1$\AA .
From Eq.(\ref{sig}), $\sigma=9.92\times10^{9}$m$^{-1}$.

Consider an ion with charge $q_{i}$, denote the distance between the extra
electron and the ion as $x$, then the static displacement caused by the extra
electron is%

\begin{equation}
d(x,q_{i})=\frac{1}{4\pi\epsilon_{0}}\frac{eq_{i}}{k_{e}\varepsilon_{0}x^{2}%
},\label{sd1}%
\end{equation}
where $k_{e}=k(r_{e})$ is spring constant of Pb-I bond in a normal crystal,
$\varepsilon_{0}$ is the static dielectric function. To get an upper limit of
$d(x,q_{i})$, let us take $\varepsilon_{0}=24.5$, $x=3.15$\AA , $q_{i}=e$,
then the largest possible static displacement of I$^{-}$ ion $d(x,q_{i}%
)=0.02$\AA $\ll\sqrt{\frac{2D}{k(r_{e})}}\ln8$. In an EP, even for the
stretched Pb-I bond, condition (\ref{str}) is still satisfied. Since $r=r_{e}$
is a minimum of $k(r)$ under condition (\ref{str}), both the compressed Pb-I
bond and the stretched Pb-I bond have larger spring constants $k(r)$, i.e.
larger vibrational frequency $\omega_{1}=\sqrt{\frac{k(r)}{m_{r}}}%
>\omega_{\text{LO}}$, which leads to a smaller vibrational entropy in a
polaron relative to that of an electron in the undeformed lattice.

\bibliographystyle{apsrev4-1}
\bibliography{ref2}

\end{document}